\documentclass[acmsmall]{acmart}

\AtBeginDocument{%
  \providecommand\BibTeX{{%
    \normalfont B\kern-0.5em{\scshape i\kern-0.25em b}\kern-0.8em\TeX}}}

\setcopyright{acmlicensed}
\acmJournal{TOSEM}
\acmYear{2023}

\usepackage{multirow}
\usepackage{arydshln}
\usepackage{subcaption}
\usepackage{colortbl}
\usepackage{amsfonts}

\begin{document}

\title{A Comprehensive Evaluation of Parameter-Efficient Fine-Tuning on Software Engineering Tasks}

\author{Wentao Zou}
    \email{DZ1832005@smail.nju.edu.cn}
\author{Qi Li}
    \email{MG21320006@smail.nju.edu.cn}
\author{Jidong Ge}
    \authornote{Corresponding Authors.}
    \orcid{0000-0003-1773-0942}
    \email{gjd@nju.edu.cn}
\author{Chuanyi Li}
    \authornotemark[1]
    \orcid{0000-0001-9270-5072}
    \email{lcy@nju.edu.cn}
    \affiliation{
      \institution{National Key Laboratory for Novel Software Technology at Nanjing University}
      \city{Nanjing}
      \state{Jiangsu}
      \country{China}
    }

\author{Xiaoyu Shen}
\affiliation{%
  \institution{Eastern Institute for Advanced Study}
  \country{China}
}
\email{xyshen@eitech.edu.cn}

\author{Liguo Huang}
\affiliation{%
  \institution{Department of Computer Science, Southern Methodist University}
  \country{USA}
}
\email{lghuang@lyle.smu.edu}

\author{Bin Luo}
\orcid{0000-0002-9036-0063}
\email{luobin@nju.edu.cn}
\affiliation{
  \institution{National Key Laboratory for Novel Software Technology at Nanjing University}
  \city{Nanjing}
  \state{Jiangsu}
  \country{China}
}

\renewcommand{\shortauthors}{Zou, et al.}

\begin{abstract}
Pre-trained models (PTMs) have achieved great success in various Software Engineering (SE) downstream tasks following the ``pre-train then fine-tune'' paradigm. As fully fine-tuning all parameters of PTMs can be computationally expensive, a widely used solution is parameter-efficient fine-tuning (PEFT), which freezes PTMs while introducing extra parameters. Though work has been done to test PEFT methods in the SE field, a comprehensive evaluation is still lacking. This paper aims to fill in this gap by evaluating the effectiveness of five PEFT methods on eight PTMs and four SE downstream tasks. For different tasks and PEFT methods, we seek answers to the following research questions: 1) Is it more effective to use PTMs trained specifically on source code, or is it sufficient to use PTMs trained on natural language text? 2) What is the impact of varying model sizes? 3) How does the model architecture affect the performance? Besides effectiveness, we also discuss the efficiency of PEFT methods, concerning the costs of required training time and GPU resource consumption. We hope that our findings can provide a deeper understanding of PEFT methods on various PTMs and SE downstream tasks. All the codes and data are available at \url{https://github.com/zwtnju/PEFT.git}.
\end{abstract}



\maketitle

\section{Introduction}
Pre-training language models on large amounts of data with self-supervision, and then fine-tuning them on downstream tasks has become a dominant approach in natural language processing (NLP)~\cite{shen2017estimation,devlin2018bert,xu2019bert,mogadala2020integrating,su2020moviechats,raffel2020exploring,adelani2022few}. The same paradigm has also been applied to software engineering (SE) downstream tasks and demonstrated exceptional performances~\cite{iyer2018mapping, xu2019bert, haque2020improved, liu2021continual,song2022multi, zeng2022extensive,zhong2022standup4npr}. Despite the remarkable success, full fine-tuning is known to be cost inefficient~\cite{tang2021ast,tang2022ast,shi2023towards}, which requires fine-tuning and saving the full set of parameters for each downstream task~\cite{zheng2023towards}. When PTMs become large, a lot of GPU and memory resources are required and the computational cost is demanding.

\begin{figure}
\centerline{\includegraphics[width=.95\linewidth]{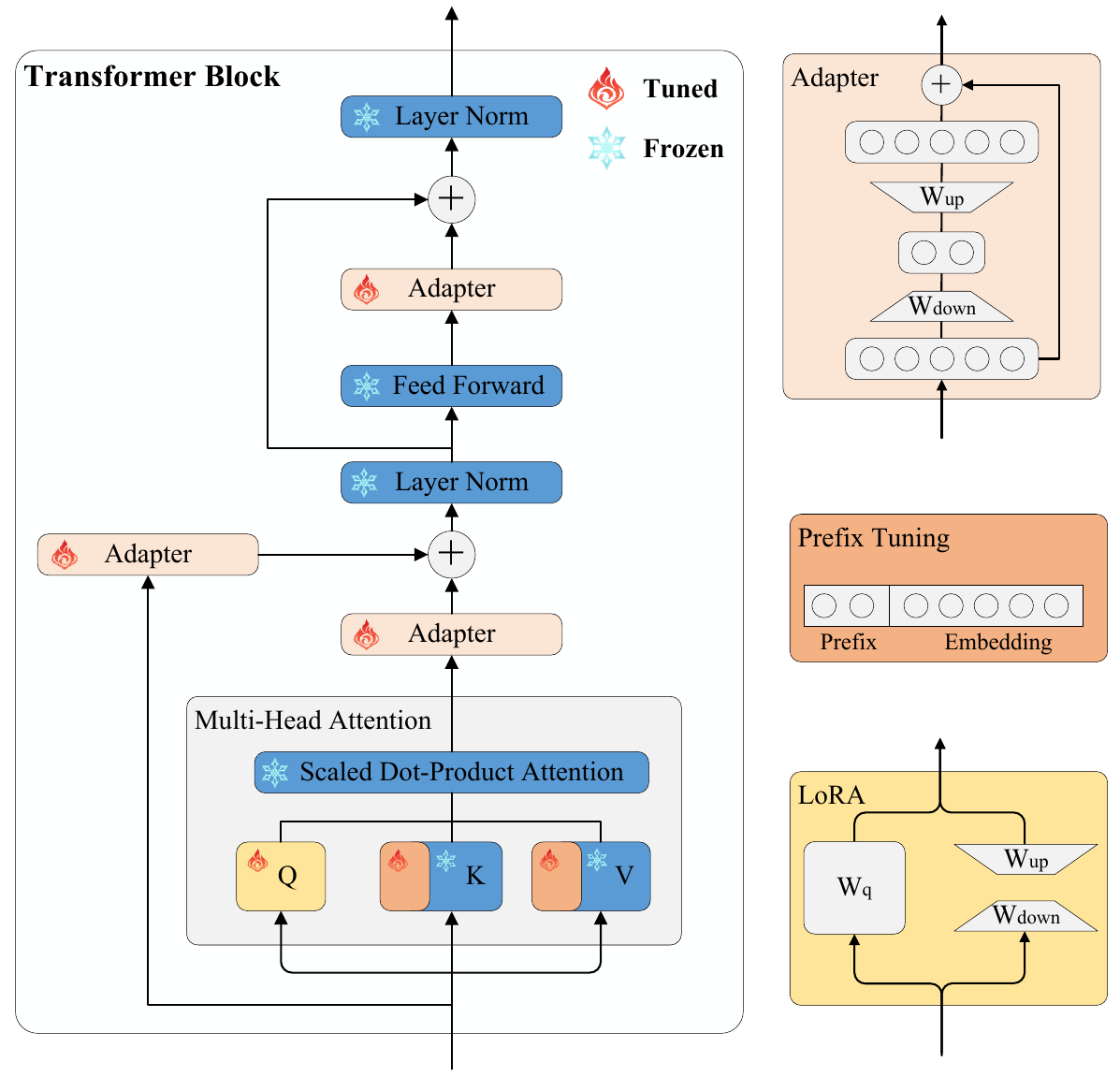}}
\caption{The Architecture of Transformer Blocks with Frozen and Fine-tuned Layers for Three Types of PEFT Methods: Adapter, Prefix Tuning, and LoRA.}
\label{fig:archi}
\end{figure}

To overcome the above drawbacks, a popular strategy is parameter-efficient fine-tuning (PEFT) \cite{fu2023effectiveness}. In contrast to full fine-tuning, PEFT only fine-tunes a limited amount of extra parameters for each task while freezing the base PTM to retain its original knowledge. Figure~\ref{fig:archi} illustrates the architecture of transformer blocks with frozen and fine-tuned layers for three types of PEFT methods: adapter~\citet{houlsby2019parameter}, prefix tuning~\cite{li2021prefix}, and LoRA~\cite{hu2021lora}. PEFT significantly reduces the computing cost as gradients flow only through the small amount of extra parameters. After training, we only need to store one fixed copy of the original large PTM and every downstream task has its own small set of extra parameters. This can also greatly reduce the storage cost. Though fine-tuning only a small set of extra, evidence has been shown that PEFT methods could even surpass the full fine-tuning approach on certain tasks~\cite{he2021towards}. 


In the SE field, various PEFT methods have been evaluated in recent years. For example, \citet{ayupov2022parameter} first tested Houlsby adapter tuning~\cite{houlsby2019parameter} and LoRA~\cite{hu2021lora} methods on four code-related tasks. They found that although PEFT approaches may achieve comparable or higher performance than full fine-tuning on code understanding tasks, they underperform full fine-tuning on code generation tasks. Later, \citet{wang2023one} used the Houlsby adapter tuning method on code search and code summarization tasks to improve the performance of PTMs and showed its effectiveness in cross-language and low-resource scenarios. Recently, \citet{liu2023empirical} conducted an empirical study of four PEFT methods in several SE scenarios, including low-resource, cross-language, and cross-project. 

Although some studies have applied PEFT to SE downstream tasks, they often focus on the effectiveness of PEFT methods on specific scenarios or tasks, lacking a comprehensive comparison of various PTMs from a holistic view. Firstly, it is also unclear whether it is necessary to develop code-specific PTMs (PTMs trained specifically on source code), or is sufficient to use text-only PTMs (PTMs trained on natural language text) and rely on the cross-modal transferability of PEFT on source code. Although PEFT methods have been well applied to cross-language and cross-project scenarios, there are few studies \cite{goel2022cross} on the cross-modal transfer ability of text-only PTMs. If text-only PTMs can achieve good results with PEFT methods, we no longer need to collect large-scale datasets to pre-train a new PTM specifically for code-related tasks. Considering the huge cost of pre-training, directly using text-only PTMs can save a lot of human and computation resources. Secondly, the size (the number of trainable parameters \cite{chatelain2022number,su2022welm}) of a PTM has a big impact on the effectiveness of fine-tuning. In general, the larger a PTM is, the better its performance will be. The purpose of PEFT is to reduce the trainable parameters to achieve close results. Therefore, it is necessary to compare the same PTM with different sizes. If the PEFT methods have close effects on different parameters, we can give priority to smaller PTMs to further reduce costs. Thirdly, \citet{niu2023empirical} have explored the performance of PTMs in different architectures, such as encoder-only or encoder-decoder architecture \cite{niu2022deep}, on various tasks. Their work highlights the effects of different architectures when fine-tuning PTMs. Different architectures of PTMs may also have different impacts on PEFT methods, which are ignored in previous studies. Guided by these aspects, we formulate the following three research questions to direct our study:

\begin{itemize}
    \item RQ1: How do text-only PTMs perform compared to code-specific PTMs of the same architecture and size?
    \item RQ2: How do the same text-only/code-specific PTMs with different sizes perform?
    \item RQ3: How do text-only/code-specific PTMs in different architectures perform?
\end{itemize}

Throughout the paper, we seek answers to the above research questions under five different PEFT methods: Houlsby~\cite{houlsby2019parameter}, Pfeiffer~\cite{pfeiffer2020mad}, Parallel~\cite{he2021towards}, Prefix~\cite{li2021prefix}, and LoRA~\cite{hu2021lora}. Experiments are run on four code-related tasks: clone detection of two code pairs, defect detection of one code snippet, code search across 6 programming languages, and code translation between Java and C\#. We include 8 PTMs in our study, including CodeBERT~\cite{feng2020codebert}, RoBERTa~\cite{liu2019roberta}, CodeT5~\cite{wang2021codet5}, T5~\cite{raffel2020exploring}, CodeT5-large~\cite{wang2021codet5}, T5-large~\cite{raffel2020exploring}, UniXcoder~\cite{guo2022unixcoder}, and BART~\cite{lewis2019bart}.

Besides the effectiveness, we also focus on the efficiency of PEFT methods. Most existing studies acknowledge that PEFT can save resources compared to full fine-tuning, but they focus on comparing the effects of PEFT and rarely present results regarding the resources required. Therefore, we further quantitatively discuss the costs of PEFT methods in terms of required training time and GPU resource consumption.

The main contributions of our paper are as follows:

\begin{itemize}

\item We show that PEFT methods are more advantageous in cross-modal transfer learning of PTMs. Compared with the results of code-specific PTMs, PEFT methods applied to text-only PTMs are more effective in keeping the performance of full fine-tuning PTMs on clone detection, defect detection, and code translation.

\item We find that compared with full fine-tuning, increasing the trainable parameters of T5 reduces the performance of the PEFT methods. CodeT5 with more trainable parameters performs better on defect detection and code translation and performs worse on clone detection and code search.

\item We present that PEFT methods applied to encoder-only architecture perform better on clone detection and defect detection than encoder-decoder architecture, while those used in T5-based encoder-decoder architecture are better on code search.

\item We investigate that PEFT PTMs obtain comparable, or even better results compared with full fine-tuning PTMs, especially on the code search and code translation. Among PEFT methods, the Parallel performs the best, which is consistent with the conclusion of~\citet{he2021towards}.

\item We demonstrate that PEFT methods can save about 10\%-30\% GPU resources compared with full fine-tuning methods. Pfeiffer needs the least GPU resources while Prefix and Houlsby require the most. PEFT does not promise to improve training speed, but it still has the potential to reduce training time. As the trainable parameters increase, fewer GPU resources and more training time are required.

\item We conduct extensive experiments and have made all the PEFT layers and PTMs used in our study publicly accessible. This allows for easy reproduction of our results and to utilize the best-performing models, providing broader applications to other software engineering scenarios.

\end{itemize}

The remainder of this paper is organized as follows. Section~\ref{sec:pre} provides some necessary prior knowledge. Section~\ref{sec:exp_setup} outlines research questions, and provides details on selected datasets and PTMs, and Section~\ref{sec:exp_rs} analyzes the results related to research questions. Section~\ref{sec:dis} discusses qualitative effectiveness analysis, quantitative efficiency analysis, and potential threats to validity. Section~\ref{sec:rw} presents related work, and Section~\ref{sec:conc} concludes this paper.

\section{Preliminaries} \label{sec:pre}
\subsection{Transformer and Pre-Trained Models}
Transformer~\cite{vaswani2017attention} is a deep learning model to solve sequence-to-sequence tasks. It can handle long text dependencies and has the flexibility to handle variable-length input and output sequences, making it suitable for tasks such as machine translation, text summarization, and sentiment analysis. Transformer has significantly influenced the NLP field as well as many other fields, such as SE.

\begin{figure}
\centerline{\includegraphics[width=.7\linewidth]{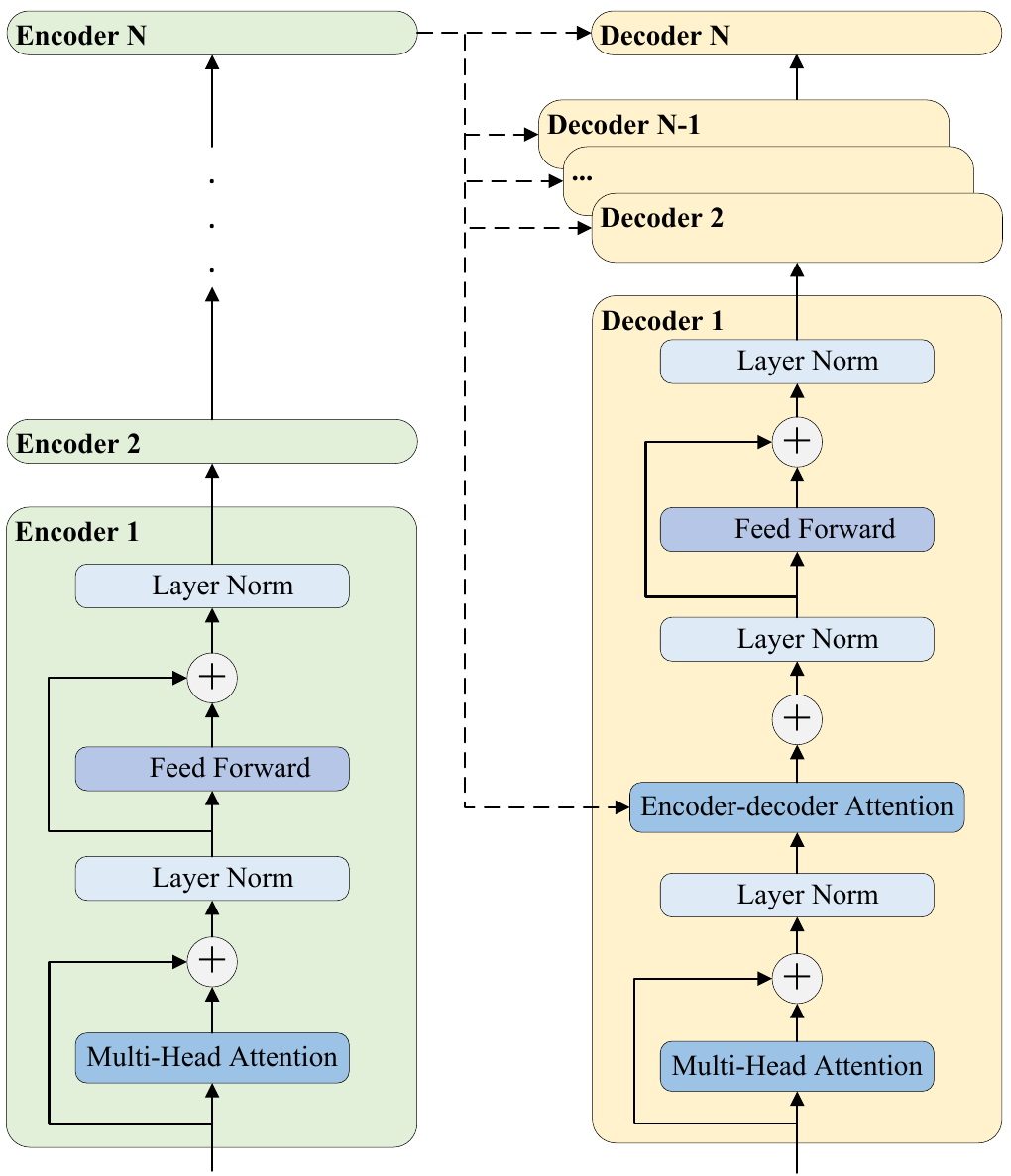}}
\caption{The simplified Transformer Architecture.}
\label{fig:transformer}
\end{figure}

The transformer has many key components, including a multi-head attention (MHA) layer, a fully connected feed-forward network (FFN), layer normalization, and residual connection. The Transformer architecture with these key components is shown in Figure~\ref{fig:transformer}.

The core of Transformer is the self-attention mechanism, which allows it to weigh the importance of different elements in a sequence. The self-attention mechanism enables the model to capture relationships between words or tokens. The calculation of attention function depends on the three components of queries $Q \in \mathbb{R}^{n \times d}$, keys $K \in \mathbb{R}^{m \times d}$ and values $V \in \mathbb{R}^{m \times d}$, where $n$ and $m$ are the number of queries and key-value pairs, and $d$ denotes the hidden size of the Transformer. The scaled dot-product attention can be formulated as follows:

\begin{equation}
    Attention(K, Q, V) = softmax(\frac{QK^T}{\sqrt{d}})V
\end{equation}

Based on self-attention, MHA performs the attention function in parallel over $N_h$ heads, where $i$th head is parameterized by $W^i_q, W^i_k, W^i_v \in \mathbb{R}^{d \times d_h}$ to project inputs to queries, keys, and values. In MHA, $d_h$ is typically set to $\frac{d}{N_h}$ to save parameters, which indicates that each attention head is operating on a lower dimensional space. Given a sequence of $m$ vectors $X \in \mathbb{R}^{m \times d}$ and a query vector $x \in X$ (the inputs of queries, keys, and values of self-attention are the same), the attention in the $i$th head can be computed as:

\begin{equation}
    head_i = Attention(xW^i_q, XW^i_k, XW^i_v) \label{eq:attention}
\end{equation}

Then, given the initial parameters $W_0 \in \mathbb{R}^{d \times d}$, the MHA concatenates the attention results of different heads as follows:

\begin{equation}
    MHA(x) = concat(head_1, \cdots, head_h)W_0
\end{equation}

FFN consists of two linear transformations with an activation function of ReLU in between, formulated as follows:
\begin{equation}
    FFN(x) = ReLU(xW_1 + b_1)W_2 + b_2
\end{equation}

where $W_1 \in \mathbb{R}^{d \times d_m}, W_2 \in \mathbb{R}^{d_m \times d} $ are trainable parameters, and $d_m$ is typically set to 4d and acts as a scaling factor.

\begin{figure}
\centerline{\includegraphics[width=1\linewidth]{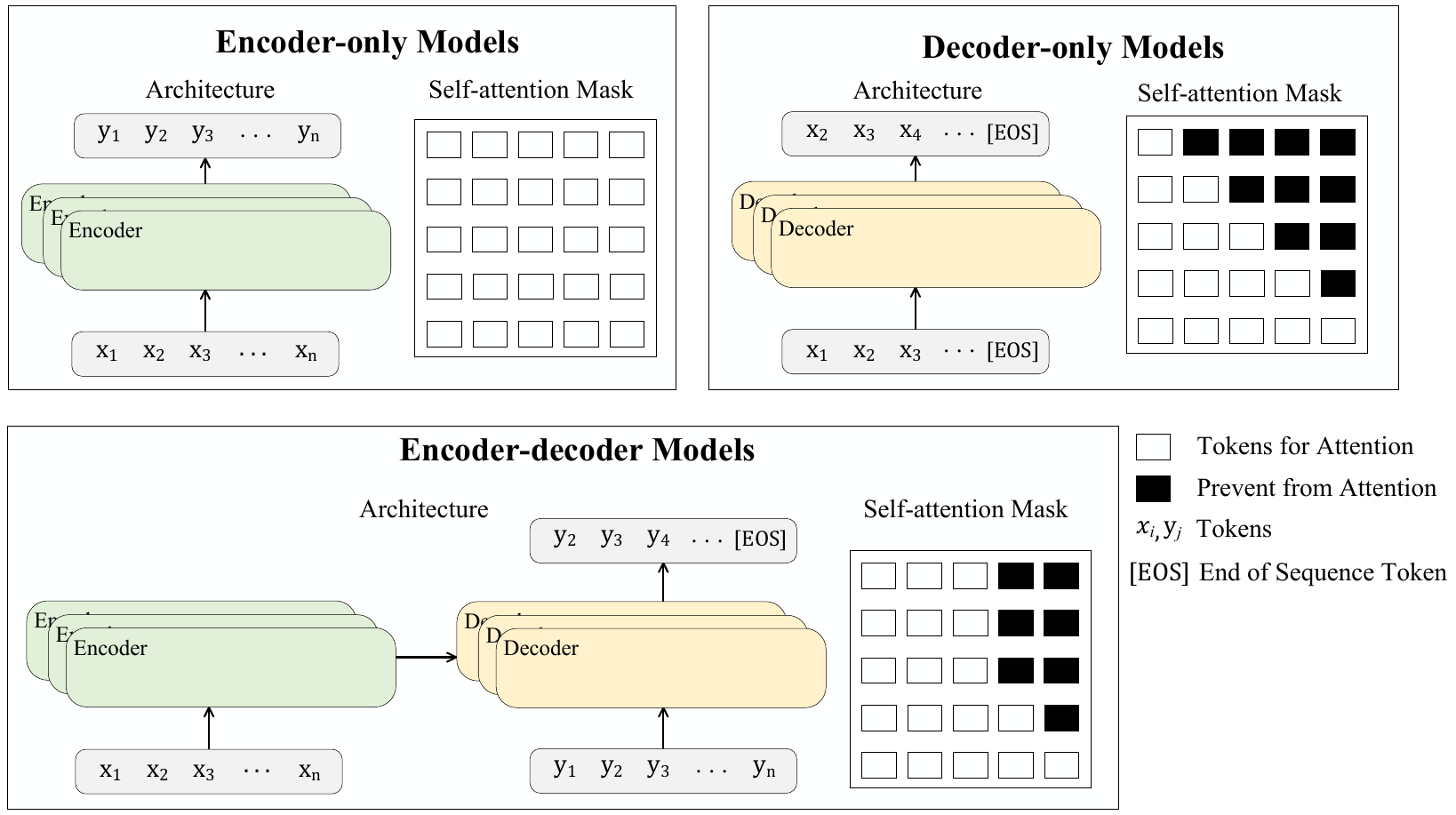}}
\caption{Different Types of Pre-trained Models.}
\label{fig:ptms}
\end{figure}

Transformer has become the foundation model and has been adapted to various models, which are called PTMs. PTMs are usually trained on extensive datasets over pre-training tasks. Pre-training tasks are specially designed general tasks that his somewhat related to downstream tasks, allowing PTMs to learn effective representations that can be applied to a wide range of downstream tasks. Transformer consists of encoders and decoders. Encoders process the input sequence, while decoders generate the output sequence based on the encoder's context and the previously generated tokens. PTMs are built upon the Transformer architecture and often use multiple encoders and/or decoders. They can be divided into encoder-only, decoder-only, and encoder-decoder models, as is shown in Figure~\ref{fig:ptms}. 

\subsection{Full Fine-Tuning and PEFT Methods}
PTMs are becoming more important because non-pre-training methods require large labeled datasets for training, which may result in slow training, information loss, and model instability. The common paradigm of using PTMs is to pre-train and then fully fine-tune, and it has proven advantages in many SE downstream tasks, such as clone detection and code search. Full fine-tuning requires tuning all parameters of a PTM, including embedding layers, MHA layers, and FFN layers, which enables the PTM to transfer learned text representations to unknown tasks. However, as PTMs become larger, the cost of full fine-tuning increases. Many techniques have been proposed to reduce the heavy computational demands of large models~\cite{gou2021knowledge, chang2021selectgen,shi2022compressing, svyatkovskiy2021fast} and a popular strategy is PEFT. PEFT focuses on optimizing fine-tuning that only tunes a small number of parameters or learns external modules for new tasks while keeping most of the PTMs' parameters frozen. PEFT can achieve similar effects with full fine-tuning and is particularly useful when computing resources and memory are limited.

\begin{figure}
\centerline{\includegraphics[width=.8\linewidth]{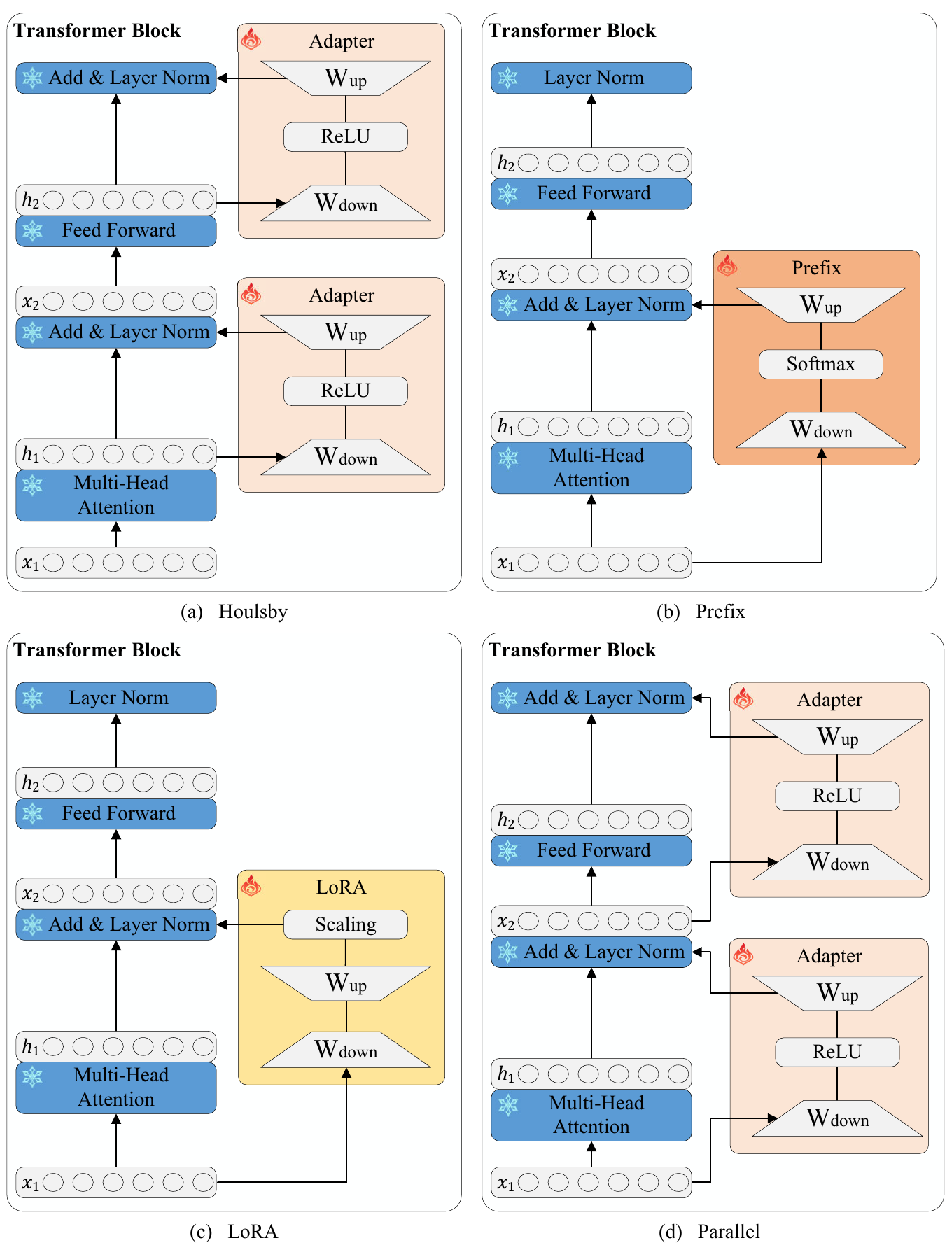}}
\caption{Illustration of used PEFT methods.}
\label{fig:peft}
\end{figure}

Next, we introduce the PEFT methods used in this paper, including Houlsby, Pfeiffer, Prefix, LoRA, and Parallel, as illustrated in Figure~\ref{fig:peft}. Houlsby and Pfeiffer use adapters in similar ways, so we only show the adapter architecture of Houlsby in the figure. Houlsby and Pfeiffer belong to ``sequential'' computation that uses $h$, the output of the MHA/FFN layer to compute $\Delta h$. In contrast, there is another way of ``parallel'' computation. The ``parallel'' methods contain Prefix, LoRA, and Parallel, and directly use $x$, the input of the MHA/FFN layer, to compute $\Delta h$.

Houlsby~\cite{houlsby2019parameter} approach inserts small modules, called adapters, after MHA and FFN layers. The adapter layer uses a down-projection $W_{down} \in \mathbb{R} ^{d \times r}$ to project the input vector to a lower-dimensional vector of dimensional size $r$, a nonlinear activation function $f$, and a up-projection $W_{up} \in \mathbb{R} ^{r \times d}$ to project lower-dimensional vector to the dimension of input vector. Given an output vector $h \in \mathbb{R} ^{d}$ of MHA/FFN layers, the output after an adapter can be represented by:

\begin{equation}
    h \leftarrow h + \Delta h, \quad \Delta h = f(hW_{down})W_{up} \label{eq:houlsby}
\end{equation}

Pfeiffer~\cite{pfeiffer2020mad} is a more efficient variant in which adapters are inserted only after the FFN layers.

Prefix~\cite{li2021prefix} introduces new parameters in the MHA layers. Specifically, it adds $l$ trainable prefix vectors to the keys and values of the MHA and concatenates prefix vectors $P_k \in \mathbb R^{l \times d} $ and $P_v \in \mathbb R^{l \times d}$ to the original keys $K$ and values $V$, respectively. Then MHA is performed on the new prefixed keys and values and the computation of $head_i$ in Eq.~\ref{eq:attention} becomes:

\begin{equation}
    head_i = Attention(xW^i_q, concat(p^i_k, XW^i_k), concat(p^i_v, XW^i_v)) \label{eq:prefix_att}
\end{equation}

where $P_k$ and $P_v$ are split into $N_h$ head vectors and $p^i_k \in \mathbb R^{l \times d/N_h}$, $p^i_v \in \mathbb R^{l \times d/N_h}$ denote the $i$th head vector of $P_k$ and $P_v$. Following~\citet{he2021towards}, the Eq.~\ref{eq:prefix_att} can be rewrote as:

\begin{equation}
    h \leftarrow (1 - \lambda (x)) h + \lambda (x) \Delta h, \quad \Delta h := softmax(xW_qP_k^\top)P_v \label{eq:prefix}
\end{equation}

where $\lambda (x)$ is a scalar representing the sum of normalized attention weights over prefixes and can be represented by:

\begin{equation}
   \lambda (x) = \frac{\sum_i exp(xW_qP_k^\top)_i}{\sum_i exp(xW_qP_k^\top)_i + \sum_j exp(xW_qW_k^\top X^\top)_j}
\end{equation}

LoRA~\cite{hu2021lora} is short for Low-Rank Adaptation. It injects trainable low-rank decomposition matrices into transformer layers. This can be applied to any weight matrix, but LoRA only applies to the query $Wq$ and value $Wv$ projection matrices of MHA layers. For any layer expressed as a matrix multiplication of the form $h=xW_0$, $x \in \mathbb{R}^d $, $W_0 \in \mathbb{R}^{d \times k}$, it performs a reparameterization such that:

\begin{equation}
    h \leftarrow h + \Delta h, \quad \Delta h = s \cdot x W_{down}W_{up} \label{eq:lora}
\end{equation}

where $s \geq 1$ is a trainable scalar hyperparameter, $W_{down} \in \mathbb{R}^{d \times r}$, $W_{up} \in \mathbb{R}^{r \times k}$ are trainable parameters, and $r$, the low-dimensional rank of the decomposition, is the most important hyperparameter.

Parallel~\cite{he2021towards} inserts adapter layers in parallel to Transformer layers, and the design of Parallel is similar to Eq.~\ref{eq:prefix} and Eq.~\ref{eq:lora}. Different from sequential adapters, the input of parallel adapters is the input of the MHA/FFN layers rather than the output of the MHA/FFN layers. According to Eq.~\ref{eq:houlsby}, given the input $x$ and the output $h$ of MHA/FFN layer, the output after Parallel can be expressed by:

\begin{equation}
        h \leftarrow h + \Delta h, \quad \Delta h = f(xW_{down})W_{up}
\end{equation}

\citet{he2021towards} also proposed two variants of parallel adapters. The multi-head parallel adapter applies the parallel adapter to modify the head attention outputs. The scaled parallel adapter transfers the composition and insertion form of LoRA into an adapter. In this paper, we use the original parallel adapters to fine-tune PTMs.

\section{Experimental Setups} \label{sec:exp_setup}
\subsection{Research Questions}
We evaluate five PEFT methods, \textbf{Houlsby, Pfeiffer, Prefix, LoRA}, and \textbf{Parallel}, and all the PEFT methods are from AdapterHub \cite{pfeiffer2020AdapterHub}. In the paper, we focus on their effectiveness on various PTMs, aiming to answer the following three research questions:

\begin{itemize}
    \item \textbf{RQ1: How do text-only PTMs perform compared to code-specific PTMs of the same architecture and size?}
    
    \cite{liu2023empirical} have studied the effectiveness of PEFT methods, but they only focus on code-specific PTMs. In addition to using code-specific PTMs on code-related tasks, another common strategy is to use PTMs pre-trained on other datasets~\cite{ma2023scope, akbar2023ethical}, such as natural language (NL). The PEFT methods have the potential for cross-modal transfer learning~\cite{weiss2016survey,zhang2022mdia,shen2022semipqa}. It is worth discussing the effectiveness of PEFT methods on text-only PTMs as they may show different effects from the code-specific PTMs. Furthermore, designing code-specific PTMs and collecting code-related datasets for pre-training is labor-intensive. If PEFT methods can help text-only PTMs achieve good results on SE downstream tasks, the required resources can be reduced.
    
    \item \textbf{RQ2: How do the same text-only/code-specific PTMs with different sizes perform?}

    The same PTM with different sizes usually have different learning abilities for code representation, and larger PTMs usually have better performance~\cite{zan2023large}. The size of a PTM is also an important factor for PEFT, as PEFT methods help PTMs with fewer parameters to achieve results that are close to or better than those with full fine-tuning. PEFT methods may close the performance gap of PTMs with different sizes, and we can use PTMs with fewer parameters to reduce the cost. Since text-only PTMs and code-specific PTMs may have inconsistent results, we discuss them with different sizes respectively.
    
    \item \textbf{RQ3: How do text-only/code-specific PTMs in different architectures perform?}

    There is evidence that different model architectures have preferences for SE downstream tasks~\cite{niu2023empirical}. For example, PTMs with encoder-only architecture are better at code understanding tasks, while PTMs with encoder-decoder architecture are better at code generation tasks. Similar conclusions may exist on PEFT methods, and we want to explore the impact of different architectures and which architecture is more suitable for PEFT methods. The same with RQ2, we discuss text-only PTMs and code-specific PTMs in different architectures respectively.
    
\end{itemize}

\subsection{Tasks, Datasets, and Evaluation Metrics}
For each research question, we fine-tune PTMs on four SE downstream tasks: \textbf{clone detection, defect detection, code search}, and  \textbf{code translation}.

Clone detection~\cite{svajlenko2014towards} is a code understanding~\cite{wang2022bridging} task that involves predicting whether pairs of code are functionally equivalent and the output is a binary classification value. We use the BigCloneBench (BCB)~\cite{lu2021codexglue} dataset, which contains over 6 million true clone pairs and 260k false clone pairs.

Defect detection~\cite{steenhoek2022empirical} is another code understanding task to predict whether a function is vulnerable. Unlike clone detection, which considers pairs of code snippets, defect detection focuses on individual code snippets. We use the Devign~\cite{zhou2019devign} dataset, which consists of functions collected from two open-source projects and contains approximately 22k training data. Compared with clone detection, the dataset of defect detection is low-resource.

Code search~\cite{yan2020code} is a code retrieval task~\cite{mou2016convolutional} to find the most relevant code snippets from a given set based on an NL query, where the NL usually refers to the source code comments. We utilize the CodeSearchNet (CSN)~\cite{husain2019codesearchnet} dataset, which includes Go, Java, JavaScript (JS), PHP, Python (Py), and Ruby languages. Low-quality queries are filtered following the UniXcoder.

Code translation~\cite{lu2021codexglue} is a code generation~\cite{yang2023deep} task that requires translating code from one programming language to another. We use the CodeTrans~\cite{lu2021codexglue} dataset, which contains parallel code snippets in Java and C\#. The PTMs need to generate C\#/Java code for the given corresponding Java/C\# code.

We give a brief introduction to datasets of these downstream tasks and the statistics of the datasets are listed in Table~\ref{dataset}. Fine-tuning uses the training and validation datasets, and the evaluation of the downstream task uses the testing datasets. These tasks are representative of various SE downstream tasks, including code classification, code retrieval, and code generation, and the datasets include high-resource, low-resource, monolingual, and multilingual. All the datasets for fine-tuning are sourced from CodeXGLUE~\cite{lu2021codexglue}, and we use their provided code to pre-process the datasets for all tasks.

For classification and retrieval tasks, metrics such as Accuracy (Acc), F1~\cite{nafi2019clcdsa}, Precision (P), Recall (R), Mean Reciprocal Rank (MRR), and Mean Average Precision (MAP) are often used. For generation tasks, metrics such as BLEU (B.)~\cite{papineni2002bleu}, and CodeBLEU (C.B.)~\cite{ren2020codebleu}, are usually used. Some generation tasks have also used variants of Accuracy, such as Exact Match (EM), indicating whether the sequence generated by the model perfectly matches the correct answer, and Computational Accuracy (CA), computing the number of times the hypothesis function generates the same output as the reference when given the same inputs. Following~\citet{niu2023empirical}, the evaluation metrics of clone detection, defect detection, and code search are F1, Acc, and MRR, separately. For code translation, since B. and C.B. cannot accurately indicate whether the predicted code and correct answer are semantically equivalent, we use EM as the primary evaluation metric and only discuss the results of EM in this paper.


\begin{table*}
\caption{A Brief Summary of Datasets. } 
\begin{center}
\begin{tabular}{c|c|c|c|c|c}
\toprule
\textbf{Task} & \textbf{Dataset Name} & \textbf{Language} & \textbf{Train Size} & \textbf{Valid Size} & \textbf{Test Size} \\ 
\midrule
Clone Detection & BigCloneBench & Java & 901,028 & 415,416 & 415,416 \\\hline
Defect Detection & Devign & C & 21,854 & 2,732 & 2,732 \\\hline
\multirow{6}{*}{Code Search} & \multirow{6}{*}{CodeSearchNet} & Go & 167,288 & 7,325 & 8,122\\
& & Java & 164,923 & 5,183 & 10,955 \\
& & JavaScript & 58,025 & 3,885 & 3,291 \\
& & PHP & 241,241 & 12,982 & 14,014 \\
& & Python & 251,820 & 13,914 & 14,918 \\
& & Ruby & 24,927 & 1,400 & 1,261 \\\hline
\multirow{2}{*}{Code Translation} & \multirow{2}{*}{CodeTrans} & Java $\rightarrow$ C\# & 10,295 & 499 & 1,000 \\
& & C\# $\rightarrow$ Java & 10,295 & 499 & 1,000 \\
\bottomrule
\end{tabular}
\label{dataset}
\end{center}
\end{table*}

\subsection{PTMs and Implementation Details}
In the paper, the used PTMs to fine-tune are \textbf{CodeBERT}, \textbf{RoBERTa}, \textbf{CodeT5}, \textbf{T5}, \textbf{CodeT5-large}, \textbf{T5-large}, \textbf{UniXcoder} and \textbf{BART}. A brief description of some important information about the used PTMs is listed in Table~\ref{ptms}. 

BERT is one of the most famous text-only PTMs, which is a bidirectional encoder-only architecture model for text representation from Transformer. It uses masked language modeling and next sentence prediction methods to train the model. Derived from BERT, RoBERTa~\cite{liu2019roberta} is an optimization encoder-only model with just the masked language modeling task and benefits from additional training, larger batch sizes, and longer sequences. BART and T5 are two classic different encoder-decoder architecture models. BART~\cite{lewis2019bart} uses denoising autoencoder tasks, such as token masking and token deletion, to construct the model. T5~\cite{raffel2020exploring} treats every task as a text-to-text problem, which involves feeding input text into the model and generating target text. CodeBERT~\cite{feng2020codebert} is one of the first encoder-only code-specific PTMs. It adapts BERT for SE downstream tasks and proposes a new replaced token detection task, training on both natural languages and programming languages. Based on T5, CodeT5~\cite{wang2021codet5} introduces identifier-aware pre-training tasks to leverage code identifiers. This strategy improves the performance on SE downstream tasks compared with T5. In addition to individual architectures, UniXcoder~\cite{guo2022unixcoder} implements a unified cross-modal architecture and uses prefixes and different self-attention masks to control the behavior of the model. It is pre-trained on multi-modal data, which includes code, comments, and AST.

\begin{table*}
\caption{A Brief Description of PTMs.} 
\begin{center}
\resizebox{1\linewidth}{!}{\begin{tabular}{c|c|c|c|c}
\toprule
\textbf{PTMs} & \textbf{Field} & \textbf{Architecture} & \textbf{Learned Languages} & \textbf{Model Size (M)} \\ 
\midrule
CodeBERT & code-specific & BERT-based encoder-only & Go, Java, JavaScript, PHP, Python, Ruby & 125 \\
RoBERTa & text-only & BERT-based encoder-only & NL text & 125 \\ \hline
CodeT5 & code-specific & T5-based encoder-decoder & Go, Java, JavaScript, PHP, Python, Ruby, C, C\# & 223 \\
T5 & text-only & T5-based encoder-decoder & NL text & 223 \\ \hline
CodeT5-large & code-specific & T5-based encoder-decoder & Go, Java, JavaScript, PHP, Python, Ruby, C, C\# & 738 \\
T5-large & text-only & T5-based encoder-decoder & NL text & 738 \\ \hline
UniXcoder & code-specific & cross-modal encoder-only & Go, Java, JavaScript, PHP, Python, Ruby & 126 \\
BART & text-only & BART-based encoder-decoder & NL text & 139 \\
\bottomrule
\end{tabular}}
\label{ptms}
\end{center}
\end{table*}

All these PTMs come from Hugging Face\footnote{\url{https://huggingface.co}}, and we select different PTMs for each research question. For RQ1, we employed two text-only PTMs (RoBERTa and T5) and code-specific PTMs (CodeBERT and CodeT5) for fine-tuning. CodeBERT and RoBERTa have the same size and are BERT-based architecture. T5 and CodeT5 share the same size and T5-based architecture. For RQ2, we selected the T5 and CodeT5 with different sizes: basic and large. We used T5 and CodeT5 to refer to the basic versions to maintain consistency with RQ1, and T5-large and CodeT5-large to refer to the large versions of T5 and CodeT5 respectively. For RQ3, we applied three code-specific PTMs (CodeBERT, UniXcoder, and CodeT5) and text-only PTMs (RoBERTa, BART, and T5). CodeBERT and UniXcoder are both encoder-only architectures, but UniXcoder can use the input prefixes to learn encoder-decoder and decoder-only architectures, so they can be regarded as different architectures. Similarly, T5 uses the end-to-end pre-training strategy, while BART uses denoising autoencoding to pre-train. For convenience, they can also be regarded as different encoder-decoder architectures. In the following sections, we refer to the architectures of CodeBERT, UniXcoder, BART, and T5 as BERT-based encoder-only, cross-modal encoder-only, BART-based encoder-decoder and T5-based encoder-decoder architecture respectively. 

\begin{table*}
\caption{Hyperparameters for Fine-Tuning Downstream Tasks.} 
\begin{center}
\resizebox{1\linewidth}{!}{\begin{tabular}{lcccc}
\toprule
\textbf{Types} & \textbf{Clone Detection} & \textbf{Defect Detection} & \textbf{Code Search} & \textbf{Code Translation} \\ 
\midrule
\textbf{Vector Dimension}: \\
\quad T5-large or CodeT5-large & 128 & 256 & code: 256; nl: 128 & input: 512; output: 512 \\
\quad Other Pre-trained Models & 400 & 400 & code: 256; nl: 128 & input: 512; output: 512 \\
\textbf{Batch Size}: \\
\quad All Pre-trained Models & 32 & 32 & 64 & 32 \\
\textbf{Training Epochs}: \\
\quad All Pre-trained Models & 2 & 5 & JavaScript: 5; Ruby: 10; Others: 2 & 50k steps \\
\textbf{Learning Rate}: \\
\quad All Pre-trained Models & 5e-5 & 2e-5 & 2e-5 & 5e-5 \\
\bottomrule
\end{tabular}}
\label{param}
\end{center}
\end{table*}


Our implementation for fine-tuning is derived from two open-source repositories. We adopt code from the UniXcoder\footnote{\url{https://github.com/microsoft/CodeBERT/tree/master/UniXcoder}} repository on the code search task and CodeXGLUE\footnote{\url{https://github.com/microsoft/CodeXGLUE}} repository on the other three tasks. We use their original code directly for fine-tuning encoder-only models and modify the original code to accommodate encoder-decoder models. The key hyperparameters for PTMs to fine-tune on different downstream tasks are listed in Table~\ref{param} and other hyperparameters are set to follow their original release. All experiments rely on pure Python packages and are conducted on servers with 2/4 GPUs of NVIDIA Tesla V100 32GB.

\section{Experimental Results} \label{sec:exp_rs}
We list and analyze the experimental results for all research questions in this section.

\subsection{RQ1: Code-specific PTMs and Text-only PTMs of Same Size and Architecture}

\textit{Clone Detection:} From Table~\ref{rq1_cd_dd_ct_bert} and Table~\ref{rq1_cd_dd_ct_t5}, Parallel outperforms other PEFT methods on all the four PTMs, no matter the code-specific or text-only PTMs. Compared with CodeBERT, PEFT methods applied to RoBERTa are more effective in keeping the performance of PTMs. The results on the best PEFT and full fine-tuning also show that PEFT methods applied to T5 are more effective than CodeT5. Thus, PEFT methods are more useful for text-only PTMs than code-specific PTMs on clone detection.


\begin{table*}
\caption{[RQ1] The Results of BERT-based PTMs on Clone Detection, Defect Detection, and Code Translation. The subscript \textbf{Full} represents the result of full fine-tuning, \textbf{Best} indicates the result of the best PEFT method, and other subscripts mean the results of the corresponding PEFT methods.}
\begin{center}
\begin{tabular}{c|c|c|cc}
\toprule
\multirow{3}{*}{Fine-Tune Methods} & Clone Detection & Defect Detection & \multicolumn{2}{c}{Code Translation} \\ 
\cline{2-5} & BCB & Devign & Java $\rightarrow$ C\# & C\# $\rightarrow$ Java \\
\cline{2-5} & F1 & Acc & EM & EM \\
\midrule
$CodeBERT_{Full}$ & 94.95 & 63.91 & 59 & 60.1 \\ 
\cdashline{0-4}[1pt/1pt]
$CodeBERT_{Best}$ & -1.88 & -3.08 & -1.7 & +0.3 \\
$CodeBERT_{Houlsby}$ & 91.9 & 57.87 & 56.7 & 59.1 \\
$CodeBERT_{Pfeiffer}$ & 91.62 & 56.41 & 56.4 & 58.3 \\
$CodeBERT_{Parallel}$ & \textbf{93.07} & \textbf{60.83} & \textbf{57.3} & \textbf{60.4} \\
$CodeBERT_{Prefix}$ & 89.96 & 56.3 & 55.2 & 59 \\
$CodeBERT_{LoRA}$ & \underline{88.38} & \underline{55.6} & \underline{54.4} & \underline{57.2} \\ \hline \hline

$RoBERTa_{Full}$ & 93.5 & 56.62 & 58.8 & 60.8 \\ 
\cdashline{0-4}[1pt/1pt]
$RoBERTa_{Best}$ & -0.15 & +2.41 & -0.2 & -0.2 \\
$RoBERTa_{Houlsby}$ & 90.19 & \underline{54.03} & 56 & 59.4 \\
$RoBERTa_{Pfeiffer}$ & 90.24 & \underline{54.03} & 57.4 & 58.8 \\
$RoBERTa_{Parallel}$ & \textbf{93.35} & \textbf{58.93} & \textbf{58.6} & \textbf{60.6} \\
$RoBERTa_{Prefix}$ & 89.63 & 55.53 & 56.1 & \underline{58.3} \\
$RoBERTa_{LoRA}$ & \underline{87.37} & 54.06 & \underline{55} & 58.4 \\ 
\bottomrule
\end{tabular}
\label{rq1_cd_dd_ct_bert}
\end{center}
\end{table*}

\begin{table*}
\caption{[RQ1] The Results of T5-based PTMs on Clone Detection, Defect Detection, and Code Translation.}
\begin{center}
\begin{tabular}{c|c|c|cc}
\toprule
\multirow{3}{*}{Fine-Tune Methods} & Clone Detection & Defect Detection & \multicolumn{2}{c}{Code Translation} \\ 
\cline{2-5} & BCB & Devign & Java $\rightarrow$ C\# & C\# $\rightarrow$ Java \\
\cline{2-5} & F1 & Acc & EM & EM \\
\midrule
$CodeT5_{Full}$ & 93.93 & 63.65 & 65.6 & 67.3 \\
\cdashline{0-4}[1pt/1pt]
$CodeT5_{Best}$ & -1.2 & -6.4 & -5.4 & -1.5 \\
$CodeT5_{Houlsby}$ & 90.41 & 56.55 & 49.7 & 55 \\
$CodeT5_{Pfeiffer}$ & 92.67 & 56.37 & 52 & 57.3 \\
$CodeT5_{Parallel}$ & \textbf{92.73} & \textbf{57.25} & \textbf{60.2} & \textbf{65.8} \\
$CodeT5_{Prefix}$ & 90.81 & 55.53 & 44.3 & \underline{47.8} \\
$CodeT5_{LoRA}$ & \underline{88.28} & \underline{54.5} & \underline{42.8} & 52 \\ \hline \hline

$T5_{Full}$ & 91.86 & 58.42 & 52 & 52.2 \\ 
\cdashline{0-4}[1pt/1pt]
$T5_{Best}$ & -1.24 & +2.3 & +2.3 & +3 \\
$T5_{Houlsby}$ & 88.57 & 54.06 & 37.2 & 43.8 \\
$T5_{Pfeiffer}$ & 90.2 & \underline{53.99} & 41.3 & 45.3 \\ 
$T5_{Parallel}$ & \textbf{90.62} & \textbf{60.72} & \textbf{54.3} & \textbf{55.2} \\
$T5_{Prefix}$ & 82.78 & 54.06 & 33 & 37.9 \\
$T5_{LoRA}$ & \underline{84.56} & 54.47 & \underline{31.2} & \underline{33.5} \\
\bottomrule
\end{tabular}
\label{rq1_cd_dd_ct_t5}
\end{center}
\end{table*}

\begin{table*}
\caption{[RQ1] MRR Score (\%) of BERT-based PTMs on Code Search.}
\begin{center}
\begin{tabular}{c|cccccc}
\toprule
Fine-Tune Methods & Go & Java & JavaScript & PHP & Python & Ruby \\
\midrule
$CodeBERT_{Full}$ & 90.4 & 70.5 & 63.3 & 64.2 & 69 & 69.6 \\ 
\cdashline{0-6}[1pt/1pt]
$CodeBERT_{Best}$ & -0.5 & -0.7 & -0.9 & -0.3 & -0.5 & -1.1 \\
$CodeBERT_{Houlsby}$ & 86.7 & 64.4 & 49.6 & 59.7 & 63.3 & 51.6 \\
$CodeBERT_{Pfeiffer}$ & 86.6 & 63.9 & 49.9 & 59.3 & 62.8 & 51.4 \\
$CodeBERT_{Parallel}$ & \textbf{89.9} & \textbf{69.8} & \textbf{62.4} & \textbf{63.9} & \textbf{68.5} & \textbf{68.5} \\
$CodeBERT_{Prefix}$ & 87.4 & 65.2 & 54.9 & 59.4 & 63 & 56.2 \\
$CodeBERT_{LoRA}$ & \underline{84}   & \underline{55.8} & \underline{46.5} & \underline{53.9} & \underline{57.5} & \underline{41.2} \\ \hline \hline

$RoBERTa_{Full}$ & 88.5 & 64.5 & 54.6 & 59.8 & 62.2 & 60.6 \\ 
\cdashline{0-6}[1pt/1pt]
$RoBERTa_{Best}$ & -0.9 & -3.7 & -3.8 & -3.3 & -3.8 & -3.4 \\
$RoBERTa_{Houlsby}$ & 81.9 & 49.3 & 36.1 & 47.1 & 48   & 35.2 \\
$RoBERTa_{Pfeiffer}$ & 82.3 & 49.7 & 37.9 & 47.1 & 48.5 & 41.2 \\
$RoBERTa_{Parallel}$ & \textbf{87.6} & \textbf{60.8} & \textbf{50.8} & \textbf{56.5} & \textbf{58.4} & \textbf{57.2} \\
$RoBERTa_{Prefix}$ & 83.6 & 50.7 & 39.5 & 48.1 & 48.2 & 45.4 \\
$RoBERTa_{LoRA}$ & \underline{78.5} & \underline{43.3} & \underline{29.4} & \underline{42.2} & \underline{43.3} & \underline{33.8} \\
\bottomrule
\end{tabular}
\label{rq1_cs_bert}
\end{center}
\end{table*}

\begin{table*}
\caption{[RQ1] MRR Score (\%) of T5-based PTMs on Code Search.}
\begin{center}
\begin{tabular}{c|cccccc}
\toprule
Fine-Tune Methods & Go & Java & JavaScript & PHP & Python & Ruby \\
\midrule
$CodeT5_{Full}$ & 91 & 72.3 & 68 & 65.3 & 70.5 & 73.2 \\ 
\cdashline{0-6}[1pt/1pt]
$CodeT5_{Best}$ & -0.5 & -0.1 & +0.1 & -0.1 & +0.3 & +0.9 \\
$CodeT5_{Houlsby}$ & 88.5 & 68.8 & 62.4 & 62.7 & 67.5 & 67.7 \\
$CodeT5_{Pfeiffer}$ & 89.3 & 69.9 & 64.4 & 63.5 & 68.6 & 69.4 \\
$CodeT5_{Parallel}$ & \textbf{90.5} & \textbf{72.2}  & \textbf{68.1} & \textbf{65.2} & \textbf{70.8} & \textbf{74.1} \\
$CodeT5_{Prefix}$ & \underline{87.4} & \underline{62.2} & \underline{54.9} & \underline{58.1} & \underline{60.7} & \underline{63.3} \\
$CodeT5_{LoRA}$ & 87.5 & 67.4 & 58.2 & 61.4 & 65.9 & 63.7 \\ \hline \hline

$T5_{Full}$ & 86.1 & 57.5 & 45.8 & 53.2 & 56.3 & 49 \\ 
\cdashline{0-6}[1pt/1pt]
$T5_{Best}$ & +1.7 & +5.3 & +6.6 & +3.4 & +3.4 & +11.5 \\
$T5_{Houlsby}$ & 85.1 & 53.3 & 42.7 & 50   & 52.8 & 45.2 \\
$T5_{Pfeiffer}$ & 85.4 & 54.7 & 44.4 & 51   & 53.7 & 50.3 \\
$T5_{Parallel}$ & \textbf{87.8} & \textbf{62.8} & \textbf{52.4} & \textbf{56.6} & \textbf{59.7} & \textbf{60.5} \\
$T5_{Prefix}$  & 77   & 41.3 & 33.3 & 39.5 & 42 & 38.2 \\
$T5_{LoRA}$ & \underline{72.4} & \underline{35.4} & \underline{31.4} & \underline{36.4} & \underline{36.4} & \underline{33.9} \\
\bottomrule
\end{tabular}
\label{rq1_cs_t5}
\end{center}
\end{table*}

\textit{Defect Detection:} From Table~\ref{rq1_cd_dd_ct_bert} and Table~\ref{rq1_cd_dd_ct_t5}, Parallel ranks first on CodeBERT and CodeT5. Furthermore, Parallel achieves the best results on RoBERTa and T5, and it outperforms full fine-tuning by 2.41\% and 2.3\%, separately. CodeBERT and CodeT5 are pre-trained on code-related datasets, and their full fine-tuning results are better than RoBERTa and T5. However, the best PEFT methods are worse on CodeBERT/CodeT5 than that on RoBERTa/T5. Thus, PEFT methods are more effective for text-only PTMs than code-specific PTMs. LoRA achieves the worst results on CodeBERT and CodeT5. Houlsby and Pfeiffer perform the least effectively on RoBERTa and T5. Compared with the results of clone detection, the performance among different fine-tuning methods is more different. This implies that for tasks that perform poorly with full fine-tuning, the results of PEFT methods will be more unstable.


\textit{Code Translation:} From Table~\ref{rq1_cd_dd_ct_bert} and Table~\ref{rq1_cd_dd_ct_t5}, Parallel outperforms other PEFT methods for all PTMs on both sub-tasks, Prefix is the worst on two scenarios and LoRA performs worst for the other scenarios. Comparing full fine-tuning and PEFT, the results of Java $\rightarrow$ C\# translation mostly mirror those of C\# $\rightarrow$ Java translation, with the difference that full fine-tuning has a better effect on CodeBERT on C\# $\rightarrow$ Java translation. It could be that CodeBERT has learned Java, and requires less trainable parameters to learn the representation on Java $\rightarrow$ C\# translation. For other scenarios, text-only PTMs all perform better than code-specific PTMs. As a generation task, code translation is more difficult and requires the PTMs to comprehend code representation. CodeBERT and CodeT5 need full fine-tuning to extend to unlearned C\# language, while PEFT of RoBERTa and T5 can achieve comparable or even better results with full fine-tuning. It implies that PEFT is better at cross-modal transfer learning than cross-language transfer learning.

\textit{Code Search:} From Table~\ref{rq1_cs_bert} and Table~\ref{rq1_cs_t5}, Parallel performs best on all languages, while LoRA performs worst. The results of CodeBERT and CodeT5 are stable for various PEFT methods, while the results of PEFT methods on RoBERTa and T5 are quite different. The PEFT method performs worse on RoBERTa compared with CodeBERT, while the results of CodeT5 and T5 show that T5 is better when using PEFT methods. Whether a PTM is trained specifically on source code does not seem to have a significant impact on the code search task. It may be that code search is a cross-modal matching task that requires learning natural language and programming language knowledge at the same time. Other factors like the model architecture of the PTM may be more important, which we will discuss in RQ3.

\textbf{Answer to RQ1:} Among PEFT methods, Parallel always performs best on code-specific and text-only PTMs. Compared with the results of code-specific PTMs, PEFT methods applied to text-only PTMs are more effective in maintaining the performance of full fine-tuning PTMs on clone detection, defect detection, and code translation. Whether a PTM is trained specifically on source code or natural language text has no impact on the performance of code search. This shows that the PEFT methods are more advantageous in cross-modal transfer learning for PTMs.

\subsection{RQ2: The Same PTMs with Different Sizes}

\begin{table*}
\caption{[RQ2] The Results of T5 and T5-large on Clone Detection, Defect Detection, and Code Translation.} 
\begin{center}
\begin{tabular}{c|c|c|cc}
\toprule
\multirow{3}{*}{Fine-Tune Methods} & Clone Detection & Defect Detection & \multicolumn{2}{c}{Code Translation} \\ 
\cline{2-5} & BCB & Devign & Java $\rightarrow$ C\# & C\# $\rightarrow$ Java \\
\cline{2-5} & F1 & Acc & EM & EM \\
\midrule
$T5_{Full}$ & 91.86 & 58.42 & 52 & 52.2 \\ 
\cdashline{0-4}[1pt/1pt]
$T5_{Best}$ & -1.24 & +2.3 & +2.3 & +3 \\
$T5_{Houlsby}$ & 88.57 & 54.06 & 37.2 & 43.8 \\
$T5_{Pfeiffer}$ & 90.2 & \underline{53.99} & 41.3 & 45.3 \\ 
$T5_{Parallel}$ & \textbf{90.62} & \textbf{60.72} & \textbf{54.3} & \textbf{55.2} \\
$T5_{Prefix}$ & 82.78 & 54.06 & 33 & 37.9 \\
$T5_{LoRA}$ & \underline{84.56} & 54.47 & \underline{31.2} & \underline{33.5} \\ \hline \hline

$T5-large_{Full}$ & 83.3 & 54.98 & 54.5 & 54.7 \\
\cdashline{0-4}[1pt/1pt]
$T5-large_{Best}$ & -4.86 & -0.84 & +1.3 & +2.4 \\
$T5-large_{Houlsby}$ & 77.44 & 54.03 & 45 & 46.3 \\
$T5-large_{Pfeiffer}$ & \textbf{78.44} & 54.06 & 47.5 & 50.3 \\
$T5-large_{Parallel}$ & 72.85 & 54.06 & \textbf{55.8} & \textbf{57.1} \\
$T5-large_{Prefix}$ & 71.56 & \textbf{54.14} & 34.2 & \underline{37.7} \\ 
$T5-large_{LoRA}$ & \underline{68.99} & \underline{53.92} & \underline{31.7} & 42.5 \\
\bottomrule
\end{tabular}
\label{rq2_cd_dd_ct_t5}
\end{center}
\end{table*}

\begin{table*}
\caption{[RQ2] The Results of CodeT5 and CodeT5-large on Clone Detection, Defect Detection and Code Translation.}
\begin{center}
\begin{tabular}{c|c|c|cc}
\toprule
\multirow{3}{*}{Fine-Tune Methods} & Clone Detection & Defect Detection & \multicolumn{2}{c}{Code Translation} \\ 
\cline{2-5} & BCB & Devign & Java $\rightarrow$ C\# & C\# $\rightarrow$ Java \\
\cline{2-5} & F1 & Acc & EM & EM \\
\midrule
$CodeT5_{Full}$ & 93.93 & 63.65 & 65.6 & 67.3 \\
\cdashline{0-4}[1pt/1pt]
$CodeT5_{Best}$ & -1.2 & -6.4 & -5.4 & -1.5 \\
$CodeT5_{Houlsby}$ & 90.41 & 56.55 & 49.7 & 55 \\
$CodeT5_{Pfeiffer}$ & 92.67 & 56.37 & 52 & 57.3 \\
$CodeT5_{Parallel}$ & \textbf{92.73} & \textbf{57.25} & \textbf{60.2} & \textbf{65.8} \\
$CodeT5_{Prefix}$ & 90.81 & 55.53 & 44.3 & \underline{47.8} \\
$CodeT5_{LoRA}$ & \underline{88.28} & \underline{54.5} & \underline{42.8} & 52 \\ \hline \hline

$CodeT5-large_{Full}$ & 90.65 & 61.27 & 64.4 & 67.1 \\ 
\cdashline{0-4}[1pt/1pt]
$CodeT5-large_{Best}$ & -1.97 & -2.08 & -2.2 & -1.8 \\
$CodeT5-large_{Houlsby}$ & 82.67 & 55.42 & 50.2 & 57.7 \\
$CodeT5-large_{Pfeiffer}$ & 87.68 & \underline{54.72} & 55.7 & 60.3 \\ 
$CodeT5-large_{Parallel}$ & \textbf{88.68} & \textbf{59.19} & \textbf{62.2} & \textbf{65.3} \\
$CodeT5-large_{Prefix}$ & 84.72 & 55.75 & 51.2 & 53.7 \\
$CodeT5-large_{LoRA}$ & \underline{80.77} & 55.16 & \underline{45.2} & \underline{52.2} \\
\bottomrule
\end{tabular}
\label{rq2_cd_dd_ct_codet5}
\end{center}
\end{table*}

\begin{table*}
\caption{[RQ2] MRR Score (\%) of T5 and T5-large on Code Search.}
\begin{center}
\begin{tabular}{c|cccccc}
\toprule
Fine-Tune Methods & Go & Java & JavaScript & PHP & Python & Ruby \\
\midrule
$T5_{Full}$ & 86.1 & 57.5 & 45.8 & 53.2 & 56.3 & 49 \\ 
\cdashline{0-6}[1pt/1pt]
$T5_{Best}$ & +1.7 & +5.3 & +6.6 & +3.4 & +3.4 & +11.5 \\
$T5_{Houlsby}$ & 85.1 & 53.3 & 42.7 & 50   & 52.8 & 45.2 \\
$T5_{Pfeiffer}$ & 85.4 & 54.7 & 44.4 & 51   & 53.7 & 50.3 \\
$T5_{Parallel}$ & \textbf{87.8} & \textbf{62.8} & \textbf{52.4} & \textbf{56.6} & \textbf{59.7} & \textbf{60.5} \\
$T5_{Prefix}$  & 77   & 41.3 & 33.3 & 39.5 & 42 & 38.2 \\
$T5_{LoRA}$ & \underline{72.4} & \underline{35.4} & \underline{31.4} & \underline{36.4} & \underline{36.4} & \underline{33.9} \\ \hline \hline

$T5-large_{Full}$ & 86.2 & 62.4 & 50.5 & 56.7 & 59.4 & 53.2 \\ 
\cdashline{0-6}[1pt/1pt]
$T5-large_{Best}$ & +0.2 & +2.2 & +4.4 & +1.3 & +2.3 & +11.7 \\
$T5-large_{Houlsby}$ & 85.7 & 59.8 & 47.7 & 54.5 & 57.2 & 51 \\
$T5-large_{Pfeiffer}$ & 86.2 & 61.2 & 51.1 & 55.5 & 58.6 & 57.9  \\
$T5-large_{Parallel}$  & \textbf{86.4} & \textbf{64.6} & \textbf{54.9} & \textbf{58} & \textbf{61.7} & \textbf{64.9} \\
$T5-large_{Prefix}$ & \underline{76.3} & \underline{44.8} & 36 & \underline{43.5} & \underline{45.1} & 38.6 \\
$T5-large_{LoRA}$ & 77.4 & 46.3 & \underline{34.7} & 45.2 & 47 & \underline{32.7} \\
\bottomrule
\end{tabular}
\label{rq2_cs_t5}
\end{center}
\end{table*}

\begin{table*}
\caption{[RQ2] MRR Score (\%) of CodeT5 and CodeT5-large on Code Search.}
\begin{center}
\begin{tabular}{c|cccccc}
\toprule
Fine-Tune Methods & Go & Java & JavaScript & PHP & Python & Ruby \\
\midrule
$CodeT5_{Full}$ & 91 & 72.3 & 68 & 65.3 & 70.5 & 73.2 \\ 
\cdashline{0-6}[1pt/1pt]
$CodeT5_{Best}$ & -0.5 & -0.1 & +0.1 & -0.1 & +0.3 & +0.9 \\
$CodeT5_{Houlsby}$ & 88.5 & 68.8 & 62.4 & 62.7 & 67.5 & 67.7 \\
$CodeT5_{Pfeiffer}$ & 89.3 & 69.9 & 64.4 & 63.5 & 68.6 & 69.4 \\
$CodeT5_{Parallel}$ & \textbf{90.5} & \textbf{72.2}  & \textbf{68.1} & \textbf{65.2} & \textbf{70.8} & \textbf{74.1} \\
$CodeT5_{Prefix}$ & \underline{87.4} & \underline{62.2} & \underline{54.9} & \underline{58.1} & \underline{60.7} & \underline{63.3} \\
$CodeT5_{LoRA}$ & 87.5 & 67.4 & 58.2 & 61.4 & 65.9 & 63.7 \\ \hline \hline

$CodeT5-large_{Full}$ & 90.2 & 71.6 & 67.4 & 64.6 & 70.6 & 73.8 \\ 
\cdashline{0-6}[1pt/1pt]
$CodeT5-large_{Best}$ & -0.7 & -0.2 & -0.2 & -0.4 & -0.4 & -0.7 \\
$CodeT5-large_{Houlsby}$ & 88.1 & 68.6 & 62.8 & 62.5 & 68.1 & 67.3 \\
$CodeT5-large_{Pfeiffer}$ & 88.5 & 69.6 & 64.4 & 63.2 & 68.8 & 69.8 \\
$CodeT5-large_{Parallel}$ & \textbf{89.5} & \textbf{71.4} & \textbf{67.2} & \textbf{64.2} & \textbf{70.2} & \textbf{73.1} \\
$CodeT5-large_{Prefix}$ & \underline{86.9} & \underline{64.2} & \underline{57.1} & \underline{59.7} & \underline{63.6} & 64.5 \\
$CodeT5-large_{LoRA}$ & 87.3 & 67.3 & 60.6 & 61.7 & 66.6 & \underline{60.7} \\
\bottomrule
\end{tabular}
\label{rq2_cs_codet5}
\end{center}
\end{table*}

\textit{Clone Detection:} From Table~\ref{rq2_cd_dd_ct_t5} and Table~\ref{rq2_cd_dd_ct_codet5}, Parallel performs the best on T5, CodeT5, and CodeT5-large, and Pfeiffer performs best on T5-large. The results show that increasing trainable parameters of T5 and CodeT5 will reduce the performance of PEFT methods. Among PTMs, the performance on T5-large is the worst compared with full fine-tuning. The difference between the best PEFT and full fine-tuning is almost the same between CodeT5 and CodeT5-large, while the performance of T5-large drops a lot compared to T5. Therefore, size has a greater negative impact on T5 compared to CodeT5.

\textit{Defect Detection:} From Table~\ref{rq2_cd_dd_ct_t5} and  Table~\ref{rq2_cd_dd_ct_codet5}, Parallel can exceed the performance of full fine-tuning on T5 by up to 2.3\%. Parallel also ranks first on CodeT5 and CodeT5-large, and Prefix is the best method on T5-large. The PEFT method on CodeT5-large brings improvements over CodeT5, but the performance on T5-large is worse than T5. It shows that increasing trainable parameters hurts T5 and has a positive impact on CodeT5. Besides, T5 can perform better with PEFT than CodeT5, but T5-large underperforms CodeT5-large with PEFT. This shows that code-related pre-training tasks can help PTM perform better with more trainable parameters.

\begin{table*} 
\caption{[RQ3] The Results of code-specific PTMs with Different Architectures on Clone Detection, Defect Detection and Code Translation.}
\begin{center}
\begin{tabular}{c|c|c|cc}
\toprule
\multirow{3}{*}{Fine-Tune Methods} & Clone Detection & Defect Detection & \multicolumn{2}{c}{Code Translation} \\ 
\cline{2-5} & BCB & Devign & Java $\rightarrow$ C\# & C\# $\rightarrow$ Java \\
\cline{2-5} & F1 & Acc & EM & EM \\
\midrule
$CodeBERT_{Full}$ & 94.95 & 63.91 & 59 & 60.1 \\ 
\cdashline{0-4}[1pt/1pt]
$CodeBERT_{Best}$ & -1.88 & -3.08 & -1.7 & +0.3 \\
$CodeBERT_{Houlsby}$ & 91.9 & 57.87 & 56.7 & 59.1 \\
$CodeBERT_{Pfeiffer}$ & 91.62 & 56.41 & 56.4 & 58.3 \\
$CodeBERT_{Parallel}$ & \textbf{93.07} & \textbf{60.83} & \textbf{57.3} & \textbf{60.4} \\
$CodeBERT_{Prefix}$ & 89.96 & 56.3 & 55.2 & 59 \\
$CodeBERT_{LoRA}$ & \underline{88.38} & \underline{55.6} & \underline{54.4} & \underline{57.2} \\ \hline \hline

$UniXcoder_{Full}$ & 94.68 & 64.28 & 58 & 58 \\ 
\cdashline{0-4}[1pt/1pt]
$UniXcoder_{Best}$ & -1.06 & -3.67 & +2.5 & +5.3 \\
$UniXcoder_{Houlsby}$ & 92.35 & 59.15 & \underline{59} & \textbf{63.3} \\
$UniXcoder_{Pfeiffer}$ & 92.41 & 58.6 & 59.7 & 60.8 \\
$UniXcoder_{Parallel}$ & 93.1 & \textbf{60.61} & \textbf{60.5} & 62.6 \\
$UniXcoder_{Prefix}$ & \textbf{93.62} & \underline{58.05} & 60.4 & \underline{60.5} \\
$UniXcoder_{LoRA}$ & \underline{90.81} & 58.53 & 59.7 & 61.3 \\ \hline \hline

$CodeT5_{Full}$ & 93.93 & 63.65 & 65.6 & 67.3 \\
\cdashline{0-4}[1pt/1pt]
$CodeT5_{Best}$ & -1.2 & -6.4 & -5.4 & -1.5 \\
$CodeT5_{Houlsby}$ & 90.41 & 56.55 & 49.7 & 55 \\
$CodeT5_{Pfeiffer}$ & 92.67 & 56.37 & 52 & 57.3 \\
$CodeT5_{Parallel}$ & \textbf{92.73} & \textbf{57.25} & \textbf{60.2} & \textbf{65.8} \\
$CodeT5_{Prefix}$ & 90.81 & 55.53 & 44.3 & \underline{47.8} \\
$CodeT5_{LoRA}$ & \underline{88.28} & \underline{54.5} & \underline{42.8} & 52 \\
\bottomrule
\end{tabular}
\label{rq3_cd_dd_ct_se}
\end{center}
\end{table*}

\textit{Code Translation:} From Table~\ref{rq2_cd_dd_ct_t5} and Table~\ref{rq2_cd_dd_ct_codet5}, for all the scenarios of two sub-tasks and four PTMs, Parallel is the best PEFT method. Prefix is the worst on C\# $\rightarrow$ Java for T5-large and CodeT5, and LoRA achieves the worst for other scenarios. Parallel outperforms the full fine-tuning method for T5 and T5-large and provides comparable results for CodeT5 and CodeT5-large. Increasing the trainable parameters only improves CodeT5 on Java $\rightarrow$ C\# translation, while resulting in negative improvements on C\# $\rightarrow$ Java translation. Since the improvement is larger, we believe that increasing trainable parameters is better for CodeT5 with PEFT. On the other hand, increasing the trainable parameters brings negative improvements to T5 on both sub-tasks.

\textit{Code Search:} From Table~\ref{rq2_cs_t5} and Table~\ref{rq2_cs_codet5}, Parallel achieves the best results on all languages among the four PTMs, while LoRA is the worst method for T5 and T5-large and Prefix performs worst for CodeT5 and CodeT5-large. For T5 and T5-large, the best PEFT method outperforms the full fine-tuning method, and for some languages, the improvement is large, such as Ruby. Furthermore, the PEFT method provides comparable or even better results to the full fine-tuning method for CodeT5 and CodeT5-large. Compared with CodeT5-large, the PEFT method performs better on CodeT5, especially outperforming the full fine-tuning method in JavaScript, Python, and Ruby. As the trainable parameters increase, the PEFT method also has a negative improvement on T5, but the gap is smaller compared with CodeT5.

\textbf{Answer to RQ2:} Among PEFT methods, Parallel always performs best on different model sizes. The results of CodeT5 with different sizes are more stable than those of T5. Increasing the trainable parameters of T5 reduces the performance of the PEFT method compared to full fine-tuning. CodeT5 with more trainable parameters performs better on defect detection and code translation and performs worse on clone detection and code search.

\subsection{RQ3: PTMs in Different Architectures}

\begin{table*}
\caption{[RQ3] The Results of text-only PTMs with Different Architectures on Clone Detection, Detection, and Code Translation.}
\begin{center}
\begin{tabular}{c|c|c|cc}
\toprule
\multirow{3}{*}{Fine-Tune Methods} & Clone Detection & Defect Detection & \multicolumn{2}{c}{Code Translation} \\ 
\cline{2-5} & BCB & Devign & Java $\rightarrow$ C\# & C\# $\rightarrow$ Java \\
\cline{2-5} & F1 & Acc & EM & EM \\
\midrule
$RoBERTa_{Full}$ & 93.5 & 56.62 & 58.8 & 60.8 \\ 
\cdashline{0-4}[1pt/1pt]
$RoBERTa_{Best}$ & -0.15 & +2.41 & -0.2 & -0.2 \\
$RoBERTa_{Houlsby}$ & 90.19 & \underline{54.03} & 56 & 59.4 \\
$RoBERTa_{Pfeiffer}$ & 90.24 & \underline{54.03} & 57.4 & 58.8 \\
$RoBERTa_{Parallel}$ & \textbf{93.35} & \textbf{58.93} & \textbf{58.6} & \textbf{60.6} \\
$RoBERTa_{Prefix}$ & 89.63 & 55.53 & 56.1 & \underline{58.3} \\
$RoBERTa_{LoRA}$ & \underline{87.37} & 54.06 & \underline{55} & 58.4 \\ \hline \hline

$BART_{Full}$ & 92.72 & 61.49 & 48.8 & 53.2 \\ 
\cdashline{0-4}[1pt/1pt]
$BART_{Best}$ & -0.32 & -0.62 & -1.8 & -1.5 \\
$BART_{Houlsby}$ & 88.47 & 56.7 & 41.4 & 47.5 \\
$BART_{Pfeiffer}$ & 88.47 & 55.75 & 41.1 & 47.7 \\
$BART_{Parallel}$ & \textbf{92.4} & \textbf{60.87} & \textbf{47} & \textbf{51.7} \\
$BART_{Prefix}$ & 89.11 & \underline{55.49} & 42.1 & 48.9 \\
$BART_{LoRA}$ & \underline{87.19} & 55.89 & \underline{31.9} & \underline{39.4} \\ \hline \hline

$T5_{Full}$ & 91.86 & 58.42 & 52 & 52.2 \\ 
\cdashline{0-4}[1pt/1pt]
$T5_{Best}$ & -1.24 & +2.3 & +2.3 & +3 \\
$T5_{Houlsby}$ & 88.57 & 54.06 & 37.2 & 43.8 \\
$T5_{Pfeiffer}$ & 90.2 & \underline{53.99} & 41.3 & 45.3 \\ 
$T5_{Parallel}$ & \textbf{90.62} & \textbf{60.72} & \textbf{54.3} & \textbf{55.2} \\
$T5_{Prefix}$ & 82.78 & 54.06 & 33 & 37.9 \\
$T5_{LoRA}$ & \underline{84.56} & 54.47 & \underline{31.2} & \underline{33.5} \\
\bottomrule
\end{tabular}
\label{rq3_cd_dd_ct_nlp}
\end{center}
\end{table*}

\begin{table*}
\caption{[RQ3] MRR Score (\%) of code-specific PTMs with Different Architectures on Code Search.}
\begin{center}
\begin{tabular}{c|cccccc}
\toprule
Fine-Tune Methods & Go & Java & JavaScript & PHP & Python & Ruby \\
\midrule
$CodeBERT_{Full}$ & 90.4 & 70.5 & 63.3 & 64.2 & 69 & 69.6 \\ 
\cdashline{0-6}[1pt/1pt]
$CodeBERT_{Best}$ & -0.5 & -0.7 & -0.9 & -0.3 & -0.5 & -1.1 \\
$CodeBERT_{Houlsby}$ & 86.7 & 64.4 & 49.6 & 59.7 & 63.3 & 51.6 \\
$CodeBERT_{Pfeiffer}$ & 86.6 & 63.9 & 49.9 & 59.3 & 62.8 & 51.4 \\
$CodeBERT_{Parallel}$ & \textbf{89.9} & \textbf{69.8} & \textbf{62.4} & \textbf{63.9} & \textbf{68.5} & \textbf{68.5} \\
$CodeBERT_{Prefix}$ & 87.4 & 65.2 & 54.9 & 59.4 & 63 & 56.2 \\
$CodeBERT_{LoRA}$ & \underline{84}   & \underline{55.8} & \underline{46.5}  & \underline{53.9} & \underline{57.5} & \underline{41.2} \\ \hline \hline

$UniXcoder_{Full}$ & 91.2 & 72.3 & 59.2 & 67.1 & 73.2 & 60.9 \\ 
\cdashline{0-6}[1pt/1pt]
$UniXcoder_{Best}$ & -4 & -4.9 & -5.8 & -3.7 & -4.1 & -0.6 \\
$UniXcoder_{Houlsby}$ & 73.9 & 57.3 & 47.7 & 59   & 59.3 & 55.7 \\
$UniXcoder_{Pfeiffer}$ & 72.6 & 56.8 & 46.4 & 44   & 53.4 & 54.6 \\
$UniXcoder_{Parallel}$ & 78.3 & 61.5 & \textbf{53.4} & 56.1 & 59.8 & \textbf{60.3} \\
$UniXcoder_{Prefix}$ & \textbf{87.2} & \textbf{67.4} & 48.4 & \textbf{63.4} & \textbf{69.1} & 55.6 \\
$UniXcoder_{LoRA}$ & \underline{69.9} & \underline{54.4} & \underline{44}   & \underline{42.4} & \underline{51.6} & \underline{50} \\ \hline \hline

$CodeT5_{Full}$ & 91 & 72.3 & 68 & 65.3 & 70.5 & 73.2 \\ 
\cdashline{0-6}[1pt/1pt]
$CodeT5_{Best}$ & -0.5 & -0.1 & +0.1 & -0.1 & +0.3 & +0.9 \\
$CodeT5_{Houlsby}$ & 88.5 & 68.8 & 62.4 & 62.7 & 67.5 & 67.7 \\
$CodeT5_{Pfeiffer}$ & 89.3 & 69.9 & 64.4 & 63.5 & 68.6 & 69.4 \\
$CodeT5_{Parallel}$ & \textbf{90.5} & \textbf{72.2}  & \textbf{68.1} & \textbf{65.2} & \textbf{70.8} & \textbf{74.1} \\
$CodeT5_{Prefix}$ & \underline{87.4} & \underline{62.2} & \underline{54.9} & \underline{58.1} & \underline{60.7} & \underline{63.3} \\
$CodeT5_{LoRA}$ & 87.5 & 67.4 & 58.2 & 61.4 & 65.9 & 63.7 \\
\bottomrule
\end{tabular}
\label{rq3_cs_se}
\end{center}
\end{table*}

\begin{table*}
\caption{[RQ3] MRR Score (\%) of text-only PTMs with Different Architectures on Code Search.}
\begin{center}
\begin{tabular}{c|cccccc}
\toprule
Fine-Tune Methods & Go & Java & JavaScript & PHP & Python & Ruby \\
\midrule
$RoBERTa_{Full}$ & 88.5 & 64.5 & 54.6 & 59.8 & 62.2 & 60.6 \\ 
\cdashline{0-6}[1pt/1pt]
$RoBERTa_{Best}$ & -0.9 & -3.7 & -3.8 & -3.3 & -3.8 & -3.4 \\ 
$RoBERTa_{Houlsby}$ & 81.9 & 49.3 & 36.1 & 47.1 & 48 & 35.2 \\
$RoBERTa_{Pfeiffer}$ & 82.3 & 49.7 & 37.9 & 47.1 & 48.5 & 41.2 \\
$RoBERTa_{Parallel}$ & \textbf{87.6} & \textbf{60.8} & \textbf{50.8} & \textbf{56.5} & \textbf{58.4} & \textbf{57.2} \\
$RoBERTa_{Prefix}$ & 83.6 & 50.7 & 39.5 & 48.1 & 48.2 & 45.4 \\
$RoBERTa_{LoRA}$ & \underline{78.5} & \underline{43.3} & \underline{29.4} & \underline{42.2} & \underline{43.3} & \underline{33.8} \\ \hline \hline

$BART_{Full}$ & 88.3 & 63.7 & 53.4 & 57.8 & 60.8 & 61 \\ 
\cdashline{0-6}[1pt/1pt]
$BART_{Best}$ & -0.8 & -4.8 & -3.9 & -3.8 & -4.3 & -4.1 \\ 
$BART_{Houlsby}$ & 84.1 & 50.6 & 40.9 & 46.5 & 48.3 & 46.8 \\
$BART_{Pfeiffer}$ & 84.1 & 50.1 & 40.4 & 45.7 & 47.7 & 46.8 \\
$BART_{Parallel}$ & \textbf{87.5} & \textbf{58.9} & \textbf{49.5} & \textbf{54}  & \textbf{56.5} & \textbf{56.9} \\
$BART_{Prefix}$ & 85.7 & 52.7 & 44.9 & 48.8 & 50.5 & 52.7 \\
$BART_{LoRA}$ & \underline{81.8} & \underline{45.9} & \underline{37} & \underline{42.3} & \underline{42.3} & \underline{41.1} \\ \hline \hline

$T5_{Full}$ & 86.1 & 57.5 & 45.8 & 53.2 & 56.3 & 49 \\ 
\cdashline{0-6}[1pt/1pt]
$T5_{Best}$ & +1.7 & +5.3 & +6.6 & +3.4 & +3.4 & +11.5 \\
$T5_{Houlsby}$ & 85.1 & 53.3 & 42.7 & 50   & 52.8 & 45.2 \\
$T5_{Pfeiffer}$ & 85.4 & 54.7 & 44.4 & 51   & 53.7 & 50.3 \\
$T5_{Parallel}$ & \textbf{87.8} & \textbf{62.8} & \textbf{52.4} & \textbf{56.6} & \textbf{59.7} & \textbf{60.5} \\
$T5_{Prefix}$  & 77   & 41.3 & 33.3 & 39.5 & 42 & 38.2 \\
$T5_{LoRA}$ & \underline{72.4} & \underline{35.4} & \underline{31.4} & \underline{36.4} & \underline{36.4} & \underline{33.9} \\ 

\bottomrule
\end{tabular}
\label{rq3_cs_nlp}
\end{center}
\end{table*}

\textit{Clone Detection:} From Table \ref{rq3_cd_dd_ct_se}, Parallel performs best on CodeBERT and CodeT5, Prefix is the best on UniXcoder, while LoRA is the worst method on the three code-specific PTMs. The best PTM to maintain full fine-tuning performance when using the PEFT method is UniXcoder. It shows that a cross-modal encoder-only architecture with code, AST, and comments is helpful for BERT-based encoder-only architecture. From Table \ref{rq3_cd_dd_ct_nlp}, Parallel performs best while LoRA is the worst method on the three text-only PTMs. Similar to code-specific PTMs, RoBERTa, BERT-based encoder-only architecture, performs better with PEFT methods than BART and T5 of encoder-decoder architecture. Thus, on clone detection, no matter whether code-specific or text-only PTMs, encoder-decoder architecture performs worse with PEFT methods than encoder-decoder architecture.

\textit{Defect Detection:} From Table~\ref{rq3_cd_dd_ct_se}, Parallel performs best on the three code-specific PTMs, Prefix performs worst on UniXcoder, while LoRA is the worst on CodeBERT and CodeT5. Among all the code-specific PTMs, the PEFT method performs best on CodeBERT compared with full fine-tuning. From Table~\ref{rq3_cd_dd_ct_nlp}, Parallel performs best on the three text-only PTMs while Houlsby, Pfeiffer, and Prefix are the worst methods. Furthermore, except for Parallel, the results of other methods are close, and there is a gap with full fine-tuning. Among text-only PTMs, BART performs worst and PEFT brings negative improvements while having positive improvements on both RoBERTa and T5. Compared with clone detection, the PEFT methods have worse results in code-specific PTMs, especially on CodeT5, but perform better on text-only PTMs, such as RoBERTa and T5. The results of code-specific and text-only PTMs show that the encoder-only architecture also performs better with PEFT methods than the encoder-decoder architecture on defect detection.

\textit{Code Translation:} From Table~\ref{rq3_cd_dd_ct_se}, Parallel performs the best in most scenarios, while LoRA, Houlsby and Prefix have worst-performing scenarios. Compared with full fine-tuning, PEFT methods perform best on UniXcoder, followed by CodeBERT and CodeT5. For UniXcoder, PEFT methods outperform full fine-tuning on both sub-tasks, and the value of the best PEFT methods of UniXcoder is comparable with that of CodeT5. From Table~\ref{rq3_cd_dd_ct_nlp}, Parallel performs the best in all scenarios and LoRA is the worst in most scenarios. PEFT methods perform best on T5 and can perform better than full fine-tuning. However, the performance of PEFT methods on BART is the worst, which shows that PEFT can not perform well on BART-based encoder-decoder architecture. Comparing the results of code-specific and text-only PTMs, text-only PTMs are more stable when using PEFT methods and the impact of model architecture on the performance of PEFT methods is small.

\textit{Code Search:} From Table~\ref{rq3_cs_se}, Parallel performs best on CodeBERT and CodeT5 for all languages. LoRA is the worst method on CodeBERT and UniXcoder and Prefix performs worst on CodeT5 for all languages. For UniXcoder, there is a gap between the best PEFT method and the full fine-tuning method, with a difference of about 5\%. Prefix outperforms Parallel on Go, Java, PHP, and Python languages, but underperforms on JavaScript and Ruby. CodeBERT and UniXcoder perform differently, which shows that unified cross-modal architecture hurts PEFT methods. PEFT methods on CodeT5 can achieve comparable or even better results with full fine-tuning. From Table~\ref{rq3_cs_nlp}, Parallel performs best and LoRA is the worst on all text-only PTMs for all languages. Full fine-tuning is better than the PEFT method on RoBERTa and BART, and the gap is similar for these two PTMs in all languages. Furthermore, the best PEFT method on T5 is better than full fine-tuning. Comparing the results of code-specific and text-only PTMs, T5-based encoder-decoder architecture is better for PEFT methods on code search.

Besides, the results of PEFT methods vary greatly for different languages. For example, the PTMs on the Go language can maintain good performance regardless of the architecture or PEFT method adopted. For CodeBERT, the results of PEFT methods except Parallel on JavaScript and Ruby languages are much worse than the results of full fine-tuning. This will help us identify which components of language models are more important in designing effective PEFT for different languages.

\textbf{Answer to RQ3:} Among PEFT methods, Parallel always performs best on PTMs in different architectures. PEFT methods applied to encoder-only architecture perform better on clone detection and defect detection than encoder-decoder architecture, while those used in T5-based encoder-decoder architecture are better on code search. Architecture has no obvious impact on the performance of PEFT methods on code translation.

\section{Discussion} \label{sec:dis}
\subsection{Effectiveness: Qualitative Analysis on Various Tasks}

\begin{table*}
\caption{The Cases that PEFT Methods Perform Better than Full Fine-Tuning.}
\begin{center}
\resizebox{1\linewidth}{!}{\begin{tabular}{c|c|c|cccccc|cc}
\toprule
\multirow{2}{*}{Models} & Clone Detection & Defect Detection & \multicolumn{6}{c|}{Code Search} & \multicolumn{2}{c}{Code Translation} \\ 
\cline{2-11} & BCB & Devign & Go & Java & JS & PHP & Py & Ruby & Java $\rightarrow$ C\# & C\# $\rightarrow$ Java\\
\hline
CodeBERT & $\downarrow$ & $\downarrow$ & $\downarrow$ & $\downarrow$ & $\downarrow$ & $\downarrow$ & $\downarrow$ & $\downarrow$ & $\downarrow$ & \cellcolor{gray!30}$\uparrow$ \\
\hline
UniXcoder & $\downarrow$ & $\downarrow$ & $\downarrow$ & $\downarrow$ & $\downarrow$ & $\downarrow$ & $\downarrow$ & $\downarrow$ & \cellcolor{gray!30}$\uparrow$ & \cellcolor{gray!30}$\uparrow$ \\
\hline
RoBERTa & $\downarrow$ & \cellcolor{gray!30}$\uparrow$ & $\downarrow$ & $\downarrow$ & $\downarrow$ & $\downarrow$ & $\downarrow$ & $\downarrow$ & $\downarrow$ & $\downarrow$ \\
\hline
BART & $\downarrow$ & $\downarrow$ & $\downarrow$ & $\downarrow$ & $\downarrow$ & $\downarrow$ & $\downarrow$ & $\downarrow$ & $\downarrow$ & $\downarrow$ \\
\hline
T5 & $\downarrow$ & \cellcolor{gray!30}$\uparrow$ & \cellcolor{gray!30}$\uparrow$ & \cellcolor{gray!30}$\uparrow$ & \cellcolor{gray!30}$\uparrow$ & \cellcolor{gray!30}$\uparrow$ & \cellcolor{gray!30}$\uparrow$ & \cellcolor{gray!30}$\uparrow$ & \cellcolor{gray!30}$\uparrow$ & \cellcolor{gray!30}$\uparrow$ \\
\hline
T5-large & $\downarrow$ & $\downarrow$ & \cellcolor{gray!30}$\uparrow$ & \cellcolor{gray!30}$\uparrow$ & \cellcolor{gray!30}$\uparrow$ & \cellcolor{gray!30}$\uparrow$ & \cellcolor{gray!30}$\uparrow$ & \cellcolor{gray!30}$\uparrow$ & \cellcolor{gray!30}$\uparrow$ & \cellcolor{gray!30}$\uparrow$ \\
\hline
CodeT5 & $\downarrow$ & $\downarrow$ & $\downarrow$ & $\downarrow$ & \cellcolor{gray!30}$\uparrow$ & $\downarrow$ & \cellcolor{gray!30}$\uparrow$ & \cellcolor{gray!30}$\uparrow$ & $\downarrow$ & $\downarrow$ \\
\hline
CodeT5-large & $\downarrow$ & $\downarrow$ & $\downarrow$ & $\downarrow$ & $\downarrow$ & $\downarrow$ & $\downarrow$ & $\downarrow$ & $\downarrow$ & $\downarrow$ \\
\bottomrule
\end{tabular}}
\label{full_worse_than_peft}
\end{center}
\end{table*}

Table~\ref{full_worse_than_peft} illustrates whether PEFT methods are better than full fine-tuning methods, where an up arrow $\uparrow$ indicates PEFT method outperforms full fine-tuning method, while a down arrow $\downarrow$ means full fine-tuning method performs better. In the table, although full fine-tuning methods are generally superior, there are specific cases where PEFT methods perform better. For classification tasks (clone detection and defect detection), full fine-tuning methods are better than PEFT methods in almost all cases. As these classification tasks are simpler and do not require the model to deeply understand the code, more parameters and longer training time usually achieve better results. For the retrieval task (code search), PEFT methods can achieve better results, especially for T5-based PTMs. For the generation task (code translation), PEFT methods can perform better on T5-based PTMs, CodeBERT, and UniXcoder, than full fine-tuning methods. These results underline the potential of PEFT methods to generate efficient and effective PTMs on code search and code translation.

\begin{table*}
\caption{Frequency of Achieving the Best or Worst Results on different tasks for PEFT methods.}
\begin{center}
\resizebox{1\linewidth}{!}{\begin{tabular}{c|c|c|cccccc|cc}
\toprule
\multirow{2}{*}{Fine-Tune Methods} & Clone Detection & Defect Detection & \multicolumn{6}{c|}{Code Search} & \multicolumn{2}{c}{Code Translation} \\ 
\cline{2-11} & BCB & Devign & Go & Java & JS & PHP & Py & Ruby & Java $\rightarrow$ C\# & C\# $\rightarrow$ Java \\
\hline
Houlsby & 0 & 2 - & 0 & 0 & 0 & 0 & 0 & 0 & 1 - & 1 + \\ \hline
Pfeiffer & 1 + & 3 - & 0 & 0 & 0 & 0 & 0 & 0 & 0 & 0 \\ \hline
Parallel & 6 + & 7 + & 7 + & 7 + & 8 + & 7 + & 7 + & 8 + & 8 + & 7 + \\ \hline
\multirow{2}{*}{Prefix} & \multirow{2}{*}{1 +} & 1 + & 1 + & 1 + & \multirow{2}{*}{2 -} & 1 + & 1 + & \multirow{2}{*}{1 -} & \multirow{2}{*}{0} & \multirow{2}{*}{4 -} \\ 
& & 2 - & 3 - & 3 - &  & 3 - & 3 - & & & \\ \hline
LoRA  & 8 - & 2 - & 5 - & 5 - & 6 - & 5 - & 5 - & 7 - & 7 - & 4 - \\
\bottomrule
\end{tabular}}
\label{diff_tasks}
\end{center}
\end{table*}


To provide a more intuitive understanding of various PEFT methods, we rank them based on the frequency of achieving the best or worst results on various tasks in Table~\ref{diff_tasks}, where ``$+$'' indicates the method that achieves the best result, ``$-$'' means the method that achieves the worst result, and the corresponding number represents the number of times the best/worst result is achieved. In the table, the Parallel method performs better while LoRA performs the worst when applied to fine-tune PTMs. A PEFT method can perform better on one task but perform worse on another task. For example, Houlsby is the best method on Java $\rightarrow$ C\# translation task but the worst on C\# $\rightarrow$ Java translation, although both are translation tasks. 


\textbf{Conclusion:} PEFT methods can perform comparable, or even better, results with corresponding full fine-tuning methods. They perform better on code search and code translation tasks. Among PEFT methods, the Parallel is the best method while LoRA performs worst.

\subsection{Efficiency: Required GPU Resources and Training Time}

PEFT methods can tune fewer parameters than full fine-tuning. To give a more intuitive comparison, we first list the trainable/total parameters of all the PTMs with various fine-tuning methods in Table~\ref{trainable}. For encoder-only PTMs on code translation tasks, we add extra decoder blocks to generate code. Therefore, 48M additional parameters need to be added to their trainable/total parameters. For example, the trainable/total parameters of Houlsby on CodeBERT is 50/175M.

\begin{table*}
\caption{Trainable/Total Parameters (M) of PTMs.} 
\begin{center}
\resizebox{0.95\linewidth}{!}{\begin{tabular}{c|c|c|c|c|c}
\hline
Fine-Tune Methods & RoBERTa/CodeBERT & UniXcoder & BART & T5/CodeT5 & T5-large/CodeT5-large \\ 
\hline
Full & 125/125 & 126/126 & 139/139 & 223/223 & 738/738 \\
\cdashline{0-5}[1pt/1pt]
Houlsby & 2/127 (1.6\%) & 2/128 (1.6\%) & 2/142 (1.5\%) & 4/227 (1.7\%) & 13/751 (1.8\%) \\
Pfeiffer & 1/126 (0.7\%) & 1/127 (0.7\%) & 1/140 (0.6\%) & 2/225 (0.8\%) & 6/744 (0.9\%) \\
Parallel & 7/132 (5.4\%) & 7/133 (5.3\%) & 7/147 (4.8\%) & 14/237 (6.0\%) & 50/788 (6.4\%) \\    
Prefix & 10/135 (7.3\%) & 10/136 (7.3\%) & 15/155 (10.0\%) & 30/253 (11.7\%) & 77/815 (9.5\%) \\
LoRA  & 0.3/125 (0.2\%) & 0.3/126 (0.2\%) & 0.4/140 (0.3\%) & 1/224 (0.4\%) & 2/740 (0.3\%) \\
\bottomrule
\end{tabular}}
\label{trainable}
\end{center}
\end{table*}

\begin{figure}
\centering	
\includegraphics[width=\textwidth]{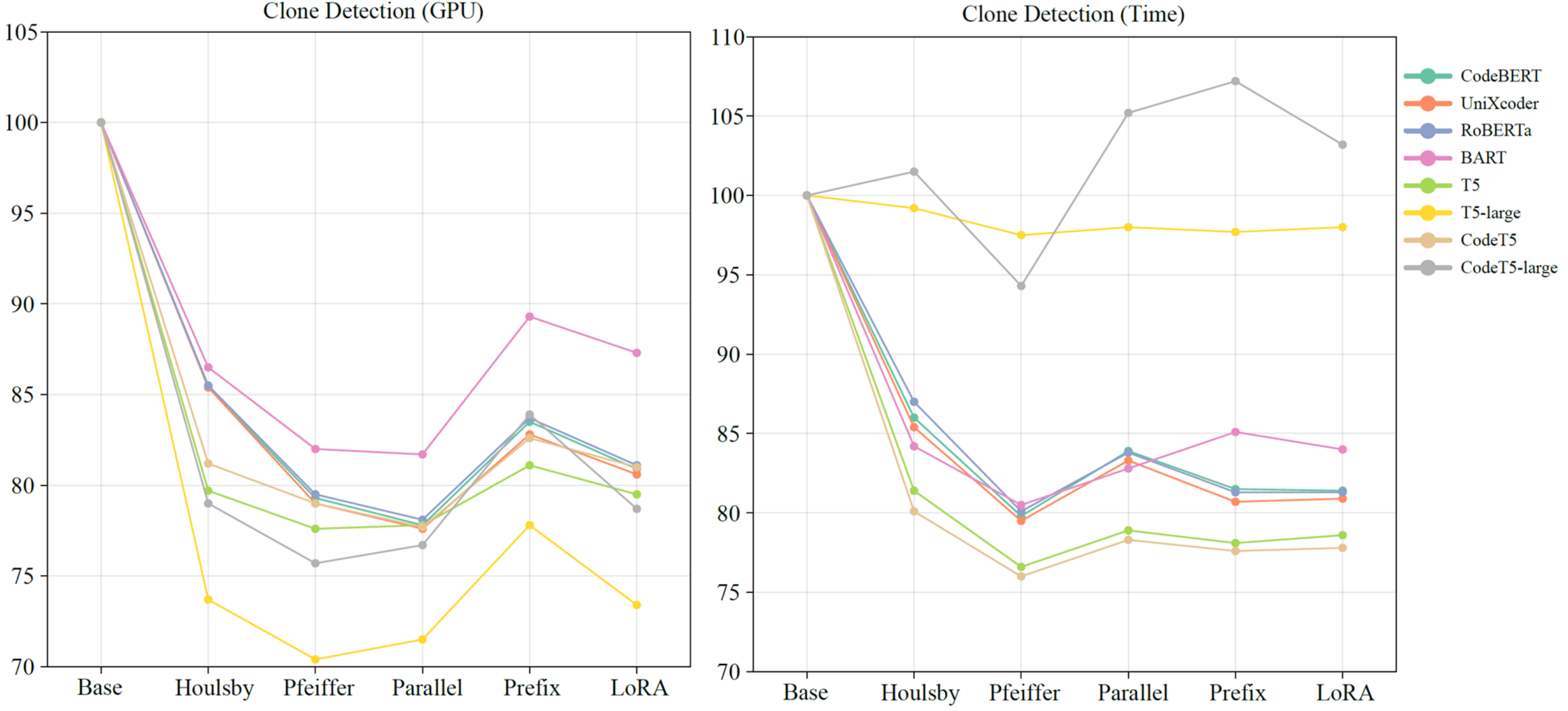}
\caption{GPU (\%) Resources and Training Time (\%) Used on Clone Detection Task.}
\label{fig:cd}
\end{figure}

\begin{figure}
\centering
\includegraphics[width=\textwidth]{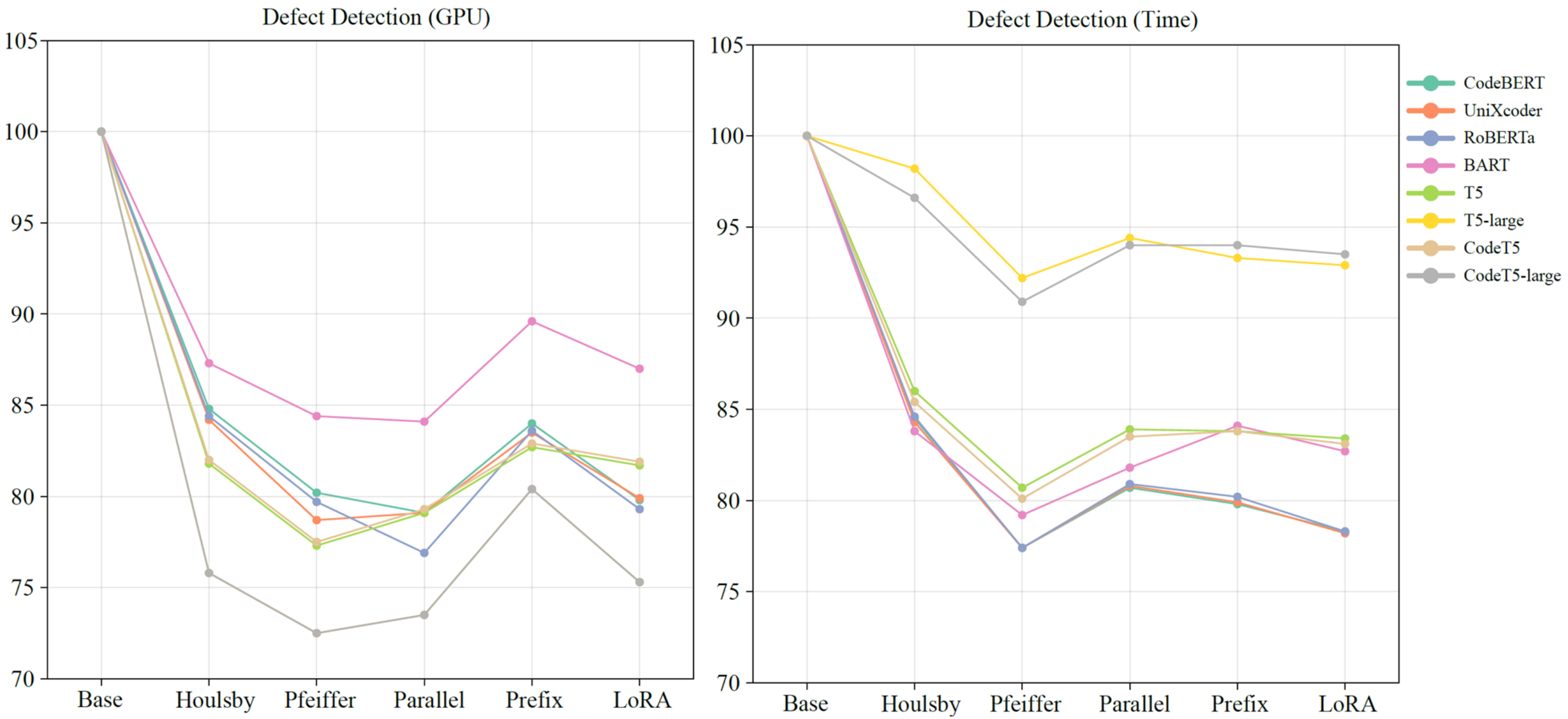}
\caption{GPU (\%) Resources and Training Time (\%) Used on Defect Detection Task.}
\label{fig:dd}
\end{figure}

\begin{figure}
\centering
\includegraphics[width=\textwidth]{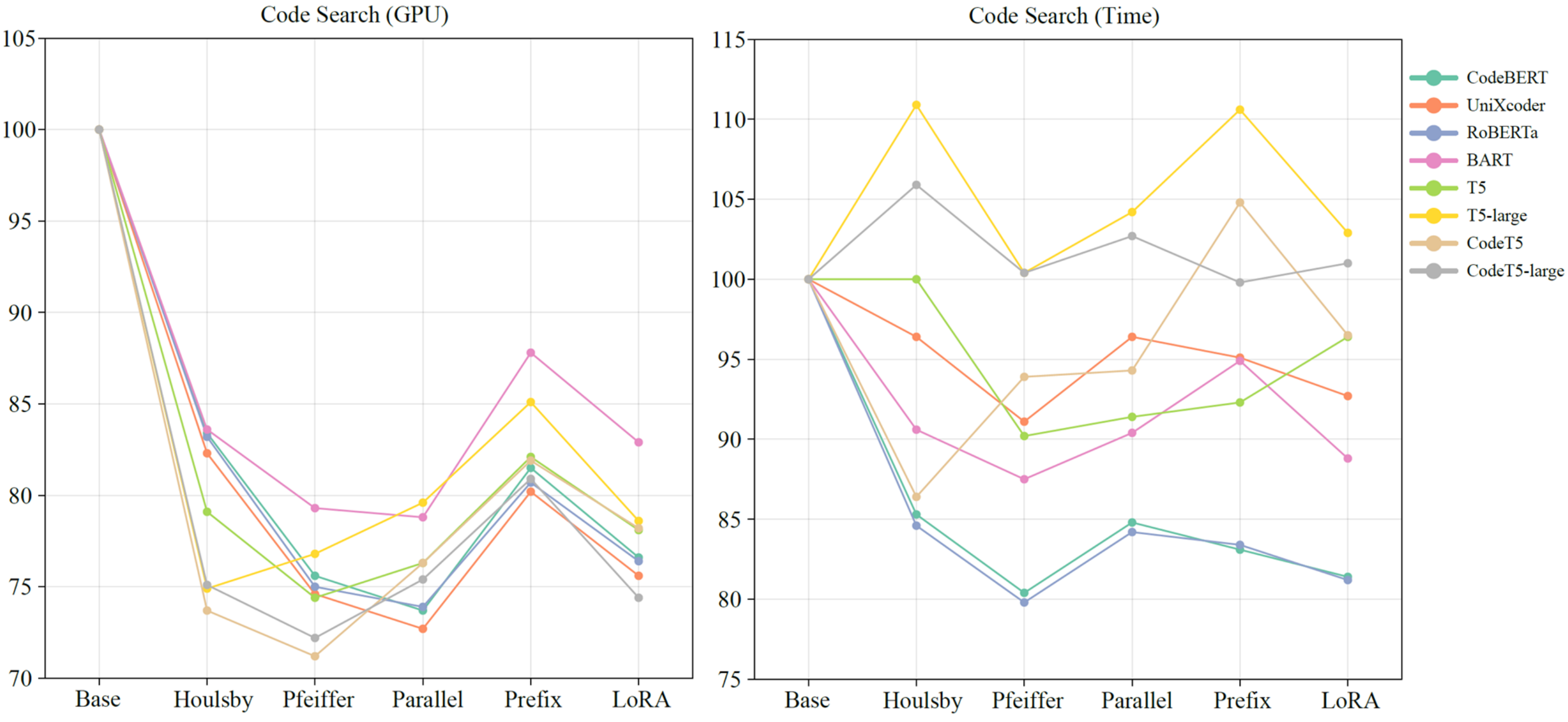}
\caption{GPU (\%) Resources and Training Time (\%) Used on JavaScript Code Search Task.}
\label{fig:cs}
\end{figure}

\begin{figure}
\centering
\includegraphics[width=\textwidth]{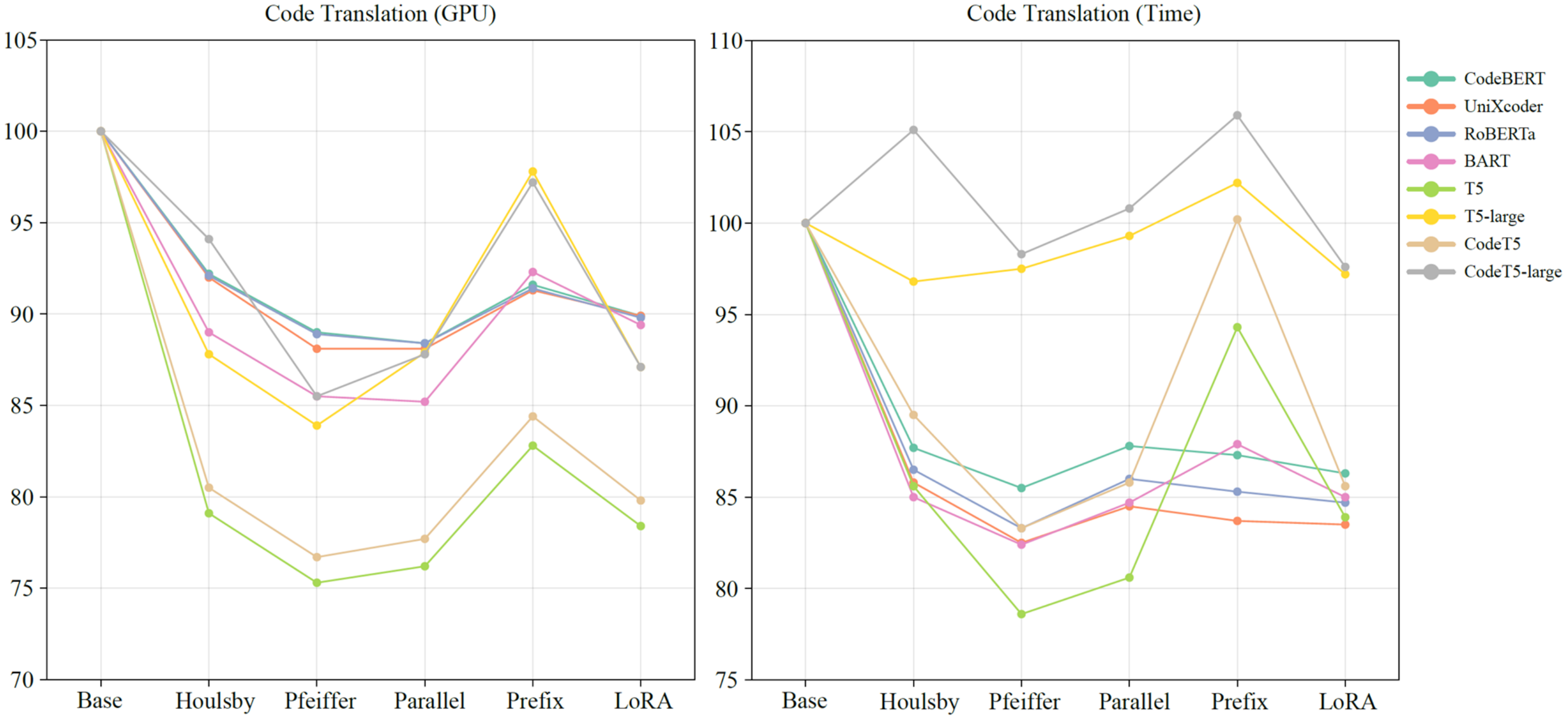}
\caption{GPU (\%) Resources and Training Time (\%) Used on Java to C\# Code Translation Task.}
\label{fig:ct}
\end{figure}

From Table~\ref{trainable}, the trainable parameters of PEFT methods are much fewer than those in the original models, but loading the models occupies GPU resources. The number of trainable parameters does not directly equate to GPU resources or its proportional occupation. Thus, we compare the required GPU resources and training time of all the PTMs with various fine-tuning methods on the four tasks. For GPU resources, we take the peak GPU occupancy of each PTM as its required GPU resources. The training time of a PTM is defined as the duration from the start of its training to its end, except for the time of data pre-processing and model evaluation. During training, we make efforts to ensure no interference from other processes on the server. Since PTMs with different parameters and architectures require inconsistent resources, to facilitate comparison, we display them uniformly as a percentage number compared to the Base. The percentage of required GPU resources and training time of four tasks are shown in Figure~{\ref{fig:cd}-\ref{fig:ct}}, corresponding to the results for each task respectively. In the figures, the ``GPU'' means required GPU resources, and ``Time'' is the actual training time. For code search, since we set the same parameters for all languages, the same method should have the same proportion of GPU resources and training time in each language. Therefore, we only illustrate the results for JavaScript language. Similarly, for code translation, we only illustrate the results of Java to C\# translation in the figures.


From the results, PEFT methods can save GPU resources compared to full fine-tuning methods, which is their primary design purpose. The used GPU resources are related to the specific implementation of PTMs and the design of PEFT methods. Generally, if a PEFT method has more trainable parameters or modifies the original model more, it takes more GPU resources. On average, the PEFT methods can save about 10\%-30\% of actual GPU resources compared to full fine-tuning. Among PEFT methods, Houlsby and Prefix take the most required GPU resources while Pfeiffer needs the least resources. On code translation tasks, LoRA also takes more GPU resources than other methods. For all the tasks, as the trainable parameters increase, fewer GPU resources are required for PEFT methods.

Although PEFT methods show clear reductions in required GPU resources, the training time of some PEFT methods is longer than the full fine-tuning method, and the results of training time are not as stable as GPU resources. For clone detection in Figure~\ref{fig:cd}, the training time of PEFT methods can exceed that of the full fine-tuning method, such as using Prefix on CodeT5-large. For defect detection in Figure~\ref{fig:dd}, all the PEFT methods are faster than full fine-tuning methods. For code search in Figure~\ref{fig:cs}, although the GPU resources show clear patterns, the training time does not follow the same clarity. A PEFT method that is fast for one PTM may be slow for another PTM. We believe that increasing the training speed is not what PEFT promises, and the complex environment (a large amount of time is spent on searching for similar code snippets) makes the training time more chaotic. The code generation task, code translation, as shown in Figure~\ref{fig:ct}, takes more time than other tasks because it requires generating code sequences other than single values. As the model grows larger, the training time of some methods, such as Prefix for T5-large and Houlsby for CodeT5-large, exceeds full fine-tuning. For all the tasks, PEFT does hold the potential to decrease the training time, and as the trainable parameters increase, more training time is required for PEFT methods.






\textbf{Conclusion:} PEFT methods can save about 10\%-30\% GPU resources compared with full fine-tuning methods. Among PEFT methods, Pfeiffer costs the least GPU resources while Prefix and Houlsby cost the most GPU resources. The increase in training speed is not what PEFT promises to improve and the training time of the PEFT method is sometimes longer than that of full fine-tuning, but it still has the potential to reduce training time. As the trainable parameters increase, fewer GPU resources and more training time are required.


\subsection{Threats to Validity} \label{sec:threat}

\textbf{Construct Validity}. As discussed in implementation details of Section~\ref{sec:exp_setup}, we reuse code from open source repositories on encoder-only models and adapt them to encoder-decoder models. While the modifications made might not result in ideal outcomes, we have tried our best to ensure the models reach their full potential.

\textbf{Internal Validity}. As a design choice, we do not perform
any hyperparameter tuning, using settings from CodeXGLUE and Table~\ref{param}. Our results are based on this set of parameters, and it is not guaranteed that the same results would be obtained under different parameters. However, for any PTM and task, the hyperparameter configurations for all PEFT methods are consistent with the corresponding full fine-tuning method, ensuring that the experimental results are obtained under fair conditions.

We are unable to deploy larger language models like CodeLlama~\cite{roziere2023code} and CodeGen~\cite{nijkamp2022codegen} due to the limitations of available resources. Smaller decoder-only models such as CodeGPT do not perform as well as other models~\cite{niu2023empirical}. Thus, our research may not provide a comprehensive view of all possible model architectures. We expect no impact of this choice on our findings.

\textbf{External Validity}. The results and findings in this work may be applicable to the specific PEFT methods, PTMs, SE downstream tasks, and corresponding datasets that we have evaluated. For other PEFT methods, PTMs, and SE downstream tasks, the same results and findings cannot be guaranteed.

\section{Related Work} \label{sec:rw}
\subsection{Code-specific Pre-Trained Models in the Software Engineering Field}
Due to the success of Transformer architecture, a large number of PTMs have been widely used in the NLP field~\cite{devlin2018bert,liu2019roberta,yang2019xlnet,radford2018improving,raffel2020exploring,lewis2019bart,su2022rocbert}. Based on these foundation models~\cite{bommasani2021opportunities}, many code-specific PTMs have been proposed in the SE field. These PTMs leverage programming language datasets for pre-training and deal with code-related downstream tasks.

CuBERT~\cite{kanade2020learning} and CodeBERT~\cite{feng2020codebert} are among the first BERT-based encoder-only models in the SE field. CuBERT employs the masked language modeling (MLM) objective, while CodeBERT proposes a new replaced token detection (RTD) objective. CuBERT is pre-trained on Python language and CodeBERT is pre-trained on six programming languages on the CodeSearchNet corpus. Later, GraphCodeBERT~\cite{guo2020graphcodebert} adds data flow information to CodeBERT and focuses on the inherent semantic-level structure of code. To make full use of data flow, it introduces two structure-aware pre-training tasks. One is to predict code structure edges, and the other is to align representations between source code and code structure. To support both code-related understanding and generation tasks, UniXcoder~\cite{guo2022unixcoder}, a unified cross-modal framework, utilizes mask attention matrices with prefix adapters to control the behavior. It is pre-trained on multi-modal data, including code, comment, and AST, and can leverage these to enhance code representation. Furthermore, it utilizes a contrastive learning task to utilize multi-modal representation and align representations among programming languages using a cross-modal generation task.

Decoder-only models~\cite{brown2020language} in the SE field are always based on GPT2~\cite{radford2019language} that utilize unidirectional language modeling objectives to predict the next token using all previous tokens. GPT-C~\cite{svyatkovskiy2020intellicode} is pre-trained in Python, C\#, JavaScript, and TypeScript programming languages to complete multilingual code. It can predict sequences of code tokens of arbitrary types, and generate up to entire lines of syntactically correct code. CodeGPT~\cite{lu2021codexglue} is pre-trained in Python and Java languages to support the code completion and text-to-code generation tasks.

Encoder-decoder models are also explored in the SE field. Based on BART, PLBART~\cite{ahmad2021unified} is pre-trained on an extensive collection of Java and Python functions and associated NL text via denoising autoencoding objectives, such as token masking, token deletion, and token infilling. Learning from T5, CodeT5~\cite{wang2021codet5} employs a unified framework to support various code-related tasks and allows for multi-task learning~\cite{zhang2018overview}. Besides, it proposes a novel identifier-aware pre-training task to leverage information of code identifiers, and a bimodal dual generation task to align code and comments. SPT-Code\cite{niu2022spt} is a sequence-to-sequence pre-trained model that uses three specially designed pre-training tasks to learn the knowledge of code, code structure, and natural language description of code.

\subsection{Parameter-Efficient Fine-Tuning Methods}
There have been over 40 different PEFT methods in various fields~\cite{rebuffi2017learning, lialin2023scaling, ding2023parameter}. Following the pioneering work of adapter tuning~\cite{houlsby2019parameter}, \citet{pfeiffer2020mad} modified adapter tuning and placed an adapter layer only after the feed-forward layer while \citet{he2021towards} placed adapter layers in parallel to the original model. Inspired by the success of prompting methods~\cite{liu2023pre}, some research focuses on additional prefix vectors. \citet{li2021prefix} used prefix tuning to add prefix vectors before input vectors and only updated prefix vectors when fine-tuning. Later, \citet{lester2021power} proposed prompt tuning that simplifies prefix tuning by only adding prompt vectors to the input sequences in the first layer and \citet{liu2023gpt} used a similar idea in the P-tuning method that employed trainable continuous prompt embeddings in concatenation with discrete prompts to increase stability. \citet{hu2021lora} proposed LoRA to inject trainable low-rank decomposition matrices into original models. To bridge this gap between the performance of PEFT and full fine-tuning, \citet{zhang2023adaptive} proposed AdaLoRA, which adaptively allocates the parameter budget among weight matrices according to their importance score. \citet{liu2022few} used $(IA)^3$ to scale inner model activations to fine-tune new parameters. BitFit~\cite{zaken2021bitfit} is a sparse fine-tuning method and only fine-tunes bias vectors. \citet{guo2020parameter} proposed diff pruning and treated fine-tuning as learning a task-specific diff vector that is applied on top of the pre-trained shared parameter vector.

In the SE field, various PEFT methods have been evaluated in recent years. \citet{ayupov2022parameter} applied PEFT methods and tested adapter tuning and LoRA on four code-related tasks, including clone detection, code summarization, code generation, and code translation. They found that PEFT can achieve comparable or higher performance than full fine-tuning on code understanding tasks, and underperform full fine-tuning on code generation tasks. Later, \citet{goel2022cross} applied adapters for cross-modal transfer to improve RoBERTa, achieving comparable performance to CodeBERT in some code-related cases. To alleviate the potentially catastrophic forgetting issue in multilingual models, \citet{wang2023one} inserted the adapters to PTMs for better results on code search and summarization tasks. They demonstrated that adapter tuning outperforms full fine-tuning in cross-lingual, multilingual, and low-resource scenarios, and adapter tuning can overcome catastrophic forgetting significantly. To fully utilize the rich syntactical information in source code, \citet{saberi2023model} proposed Named Entity Recognition (NER) adapters that can be inserted into PTMs to learn type information extracted from the AST. They inserted NER adapters in CodeBERT and they proved that adapters improve full fine-tuning on code refinement and code summarization tasks.

\subsection{Empirical Studies on Pre-Trained Models in the Software Engineering Field}
Several works have done emnlppirical studies to focus on the performance of PTMs on SE downstream tasks. \citet{chirkova2021empirical} conducted studies of the capabilities of Transformers to utilize syntactic information in different tasks. \citet{zhou2021assessing} showed the generalizability of CodeBERT to achieve higher or comparable performance than specialized solutions designed for the code search and just-in-time defect prediction tasks. \citet{ahmed2022multilingual} found that identifiers are a very important element of training data for software engineering tasks and available multilingual training data (across different languages) can be used to amplify the performance of PTMs. To better understand how monolingual and multilingual PTMs affect different programming languages, \citet{chen2022transferability} conducted and analyzed over a hundred models. \citet{steenhoek2023empirical} reproduced 9 DNN models on two vulnerability detection datasets to have a good understanding of these models. \citet{mastropaolo2023robustness} let GitHub Copilot automatically generate Java methods based on different semantically equivalent descriptions to evaluate the robustness of the code completion. Recently, \citet{tufano2023automating} focused on the impact of pre-training tasks on the performance of PTMs, and \citet{niu2023empirical} performed a comparative study of 19 PTMs on 13 SE downstream tasks to better understand these models used in source code learning. 

For PEFT methods, \citet{he2021effectiveness} studied the effectiveness of adapter-based tuning methods on several downstream NLP tasks. \citet{chen2022revisiting} conducted an empirical study of various widely used PEFT methods on NLP benchmarks. Similar research has been carried out in the SE field. \citet{wang2022no} empirically evaluated prompt tuning on CodeBERT and CodeT5 and conducted experiments on three tasks, including defect prediction, code summarization, and code translation. They found that prompt tuning outperformed fine-tuning on all three tasks and showed great potential in low-resource scenarios. This is not completely consistent with our results. We show that although PEFT performs better on the low-resource code generation task, it is not as good as full fine-tuning on the low-resource code classification task. This may be because \citet{wang2022no} did not test code classification tasks or the features of prompt tuning. Recently, \citet{liu2023empirical} conducted an empirical study of four PEFT methods on code-specific PTMs and various scenarios, including low-resource, cross-language, and cross-project scenarios. They found that in the low-resource scenario, a considerable number of samples is required for PEFT to approximate the performance of full fine-tuning. They also demonstrated that PEFT methods can improve the transfer ability of PTMs in cross-language and cross-project scenarios. Furthermore, they found that Houlsby and LoRA showed better performance than other PEFT methods, which is different from our results that Parallel is the best method. We believe that the inconsistent results are from different hyperparameters because PEFT can be affected by hyperparameters, such as the rank of LoRA~\cite{hu2021lora}. Nonetheless, our results are consistent with the conclusion of~\citet{he2021towards}, which is a comprehensive evaluation of PEFT in the NLP field in recent years.

\section{Conclusion} \label{sec:conc}
In this study, we focus on the empirical comparison of five popular PEFT and full fine-tuning methods on eight PTMs and four SE downstream tasks. Our experiments provide several important findings. For example, Parallel outperforms others among PEFT methods and PEFT methods perform better on code search and code translation. We believe our extensive experiments and evaluation can equip SE researchers with a deeper understanding of PEFT methods on various PTMs, along with their respective strengths and weaknesses. We hope that this paper will offer researchers comprehensive insights into various PEFT methods in the SE field, and inspire them to apply or design more effective PEFT methods to reduce the cost of using PTMs while maintaining their performance.

\begin{acks}
This work was supported by the National Key Research and Development Program of China (2022YFF0711404), Natural Science Foundation of Jiangsu Province, China (BK20201250), Cooperation Fund of Huawei-NJU Creative Laboratory for the Next Programming and CCF-Huawei Populus Grove Fund. 
\end{acks}

\bibliographystyle{ACM-Reference-Format}
\bibliography{sample-base}


\begin{thebibliography}{95}


\ifx \showCODEN    \undefined \def \showCODEN     #1{\unskip}     \fi
\ifx \showDOI      \undefined \def \showDOI       #1{#1}\fi
\ifx \showISBNx    \undefined \def \showISBNx     #1{\unskip}     \fi
\ifx \showISBNxiii \undefined \def \showISBNxiii  #1{\unskip}     \fi
\ifx \showISSN     \undefined \def \showISSN      #1{\unskip}     \fi
\ifx \showLCCN     \undefined \def \showLCCN      #1{\unskip}     \fi
\ifx \shownote     \undefined \def \shownote      #1{#1}          \fi
\ifx \showarticletitle \undefined \def \showarticletitle #1{#1}   \fi
\ifx \showURL      \undefined \def \showURL       {\relax}        \fi
\providecommand\bibfield[2]{#2}
\providecommand\bibinfo[2]{#2}
\providecommand\natexlab[1]{#1}
\providecommand\showeprint[2][]{arXiv:#2}

\bibitem[Adelani et~al\mbox{.}(2022)]%
        {adelani2022few}
\bibfield{author}{\bibinfo{person}{David Adelani}, \bibinfo{person}{Jesujoba Alabi}, \bibinfo{person}{Angela Fan}, \bibinfo{person}{Julia Kreutzer}, \bibinfo{person}{Xiaoyu Shen}, \bibinfo{person}{Machel Reid}, \bibinfo{person}{Dana Ruiter}, \bibinfo{person}{Dietrich Klakow}, \bibinfo{person}{Peter Nabende}, \bibinfo{person}{Ernie Chang}, {et~al\mbox{.}}} \bibinfo{year}{2022}\natexlab{}.
\newblock \showarticletitle{A Few Thousand Translations Go a Long Way! Leveraging Pre-trained Models for African News Translation}. In \bibinfo{booktitle}{\emph{Proceedings of the 2022 Conference of the North American Chapter of the Association for Computational Linguistics: Human Language Technologies}}. \bibinfo{pages}{3053--3070}.
\newblock


\bibitem[Ahmad et~al\mbox{.}(2021)]%
        {ahmad2021unified}
\bibfield{author}{\bibinfo{person}{Wasi~Uddin Ahmad}, \bibinfo{person}{Saikat Chakraborty}, \bibinfo{person}{Baishakhi Ray}, {and} \bibinfo{person}{Kai-Wei Chang}.} \bibinfo{year}{2021}\natexlab{}.
\newblock \showarticletitle{Unified pre-training for program understanding and generation}.
\newblock \bibinfo{journal}{\emph{arXiv preprint arXiv:2103.06333}} (\bibinfo{year}{2021}).
\newblock


\bibitem[Ahmed and Devanbu(2022)]%
        {ahmed2022multilingual}
\bibfield{author}{\bibinfo{person}{Toufique Ahmed} {and} \bibinfo{person}{Premkumar Devanbu}.} \bibinfo{year}{2022}\natexlab{}.
\newblock \showarticletitle{Multilingual training for software engineering}. In \bibinfo{booktitle}{\emph{Proceedings of the 44th International Conference on Software Engineering}}. \bibinfo{pages}{1443--1455}.
\newblock


\bibitem[Akbar et~al\mbox{.}(2023)]%
        {akbar2023ethical}
\bibfield{author}{\bibinfo{person}{Muhammad~Azeem Akbar}, \bibinfo{person}{Arif~Ali Khan}, {and} \bibinfo{person}{Peng Liang}.} \bibinfo{year}{2023}\natexlab{}.
\newblock \showarticletitle{Ethical Aspects of ChatGPT in Software Engineering Research}.
\newblock \bibinfo{journal}{\emph{arXiv preprint arXiv:2306.07557}} (\bibinfo{year}{2023}).
\newblock


\bibitem[Ayupov and Chirkova(2022)]%
        {ayupov2022parameter}
\bibfield{author}{\bibinfo{person}{Shamil Ayupov} {and} \bibinfo{person}{Nadezhda Chirkova}.} \bibinfo{year}{2022}\natexlab{}.
\newblock \showarticletitle{Parameter-Efficient Finetuning of Transformers for Source Code}.
\newblock \bibinfo{journal}{\emph{arXiv preprint arXiv:2212.05901}} (\bibinfo{year}{2022}).
\newblock


\bibitem[Bommasani et~al\mbox{.}(2021)]%
        {bommasani2021opportunities}
\bibfield{author}{\bibinfo{person}{Rishi Bommasani}, \bibinfo{person}{Drew~A Hudson}, \bibinfo{person}{Ehsan Adeli}, \bibinfo{person}{Russ Altman}, \bibinfo{person}{Simran Arora}, \bibinfo{person}{Sydney von Arx}, \bibinfo{person}{Michael~S Bernstein}, \bibinfo{person}{Jeannette Bohg}, \bibinfo{person}{Antoine Bosselut}, \bibinfo{person}{Emma Brunskill}, {et~al\mbox{.}}} \bibinfo{year}{2021}\natexlab{}.
\newblock \showarticletitle{On the opportunities and risks of foundation models}.
\newblock \bibinfo{journal}{\emph{arXiv preprint arXiv:2108.07258}} (\bibinfo{year}{2021}).
\newblock


\bibitem[Brown et~al\mbox{.}(2020)]%
        {brown2020language}
\bibfield{author}{\bibinfo{person}{Tom Brown}, \bibinfo{person}{Benjamin Mann}, \bibinfo{person}{Nick Ryder}, \bibinfo{person}{Melanie Subbiah}, \bibinfo{person}{Jared~D Kaplan}, \bibinfo{person}{Prafulla Dhariwal}, \bibinfo{person}{Arvind Neelakantan}, \bibinfo{person}{Pranav Shyam}, \bibinfo{person}{Girish Sastry}, \bibinfo{person}{Amanda Askell}, {et~al\mbox{.}}} \bibinfo{year}{2020}\natexlab{}.
\newblock \showarticletitle{Language models are few-shot learners}.
\newblock \bibinfo{journal}{\emph{Advances in neural information processing systems}}  \bibinfo{volume}{33} (\bibinfo{year}{2020}), \bibinfo{pages}{1877--1901}.
\newblock


\bibitem[Chang et~al\mbox{.}(2021)]%
        {chang2021selectgen}
\bibfield{author}{\bibinfo{person}{Ernie Chang}, \bibinfo{person}{Xiaoyu Shen}, \bibinfo{person}{Alex Marin}, {and} \bibinfo{person}{Vera Demberg}.} \bibinfo{year}{2021}\natexlab{}.
\newblock \showarticletitle{The SelectGen Challenge: Finding the Best Training Samples for Few-Shot Neural Text Generation}. In \bibinfo{booktitle}{\emph{Proceedings of the 14th International Conference on Natural Language Generation}}. \bibinfo{pages}{325--330}.
\newblock


\bibitem[Chatelain et~al\mbox{.}(2022)]%
        {chatelain2022number}
\bibfield{author}{\bibinfo{person}{Am{\'e}lie Chatelain}, \bibinfo{person}{Amine Djeghri}, \bibinfo{person}{Daniel Hesslow}, {and} \bibinfo{person}{Julien Launay}.} \bibinfo{year}{2022}\natexlab{}.
\newblock \showarticletitle{Is the number of trainable parameters all that actually matters?}. In \bibinfo{booktitle}{\emph{I (Still) Can't Believe It's Not Better! Workshop at NeurIPS 2021}}. PMLR, \bibinfo{pages}{27--32}.
\newblock


\bibitem[Chen et~al\mbox{.}(2022a)]%
        {chen2022transferability}
\bibfield{author}{\bibinfo{person}{Fuxiang Chen}, \bibinfo{person}{Fatemeh~H Fard}, \bibinfo{person}{David Lo}, {and} \bibinfo{person}{Timofey Bryksin}.} \bibinfo{year}{2022}\natexlab{a}.
\newblock \showarticletitle{On the transferability of pre-trained language models for low-resource programming languages}. In \bibinfo{booktitle}{\emph{Proceedings of the 30th IEEE/ACM International Conference on Program Comprehension}}. \bibinfo{pages}{401--412}.
\newblock


\bibitem[Chen et~al\mbox{.}(2022b)]%
        {chen2022revisiting}
\bibfield{author}{\bibinfo{person}{Guanzheng Chen}, \bibinfo{person}{Fangyu Liu}, \bibinfo{person}{Zaiqiao Meng}, {and} \bibinfo{person}{Shangsong Liang}.} \bibinfo{year}{2022}\natexlab{b}.
\newblock \showarticletitle{Revisiting parameter-efficient tuning: Are we really there yet?}
\newblock \bibinfo{journal}{\emph{arXiv preprint arXiv:2202.07962}} (\bibinfo{year}{2022}).
\newblock


\bibitem[Chirkova and Troshin(2021)]%
        {chirkova2021empirical}
\bibfield{author}{\bibinfo{person}{Nadezhda Chirkova} {and} \bibinfo{person}{Sergey Troshin}.} \bibinfo{year}{2021}\natexlab{}.
\newblock \showarticletitle{Empirical study of transformers for source code}. In \bibinfo{booktitle}{\emph{Proceedings of the 29th ACM joint meeting on European software engineering conference and symposium on the foundations of software engineering}}. \bibinfo{pages}{703--715}.
\newblock


\bibitem[Devlin et~al\mbox{.}(2018)]%
        {devlin2018bert}
\bibfield{author}{\bibinfo{person}{Jacob Devlin}, \bibinfo{person}{Ming-Wei Chang}, \bibinfo{person}{Kenton Lee}, {and} \bibinfo{person}{Kristina Toutanova}.} \bibinfo{year}{2018}\natexlab{}.
\newblock \showarticletitle{Bert: Pre-training of deep bidirectional transformers for language understanding}.
\newblock \bibinfo{journal}{\emph{arXiv preprint arXiv:1810.04805}} (\bibinfo{year}{2018}).
\newblock


\bibitem[Ding et~al\mbox{.}(2023)]%
        {ding2023parameter}
\bibfield{author}{\bibinfo{person}{Ning Ding}, \bibinfo{person}{Yujia Qin}, \bibinfo{person}{Guang Yang}, \bibinfo{person}{Fuchao Wei}, \bibinfo{person}{Zonghan Yang}, \bibinfo{person}{Yusheng Su}, \bibinfo{person}{Shengding Hu}, \bibinfo{person}{Yulin Chen}, \bibinfo{person}{Chi-Min Chan}, \bibinfo{person}{Weize Chen}, {et~al\mbox{.}}} \bibinfo{year}{2023}\natexlab{}.
\newblock \showarticletitle{Parameter-efficient fine-tuning of large-scale pre-trained language models}.
\newblock \bibinfo{journal}{\emph{Nature Machine Intelligence}} \bibinfo{volume}{5}, \bibinfo{number}{3} (\bibinfo{year}{2023}), \bibinfo{pages}{220--235}.
\newblock


\bibitem[Feng et~al\mbox{.}(2020)]%
        {feng2020codebert}
\bibfield{author}{\bibinfo{person}{Zhangyin Feng}, \bibinfo{person}{Daya Guo}, \bibinfo{person}{Duyu Tang}, \bibinfo{person}{Nan Duan}, \bibinfo{person}{Xiaocheng Feng}, \bibinfo{person}{Ming Gong}, \bibinfo{person}{Linjun Shou}, \bibinfo{person}{Bing Qin}, \bibinfo{person}{Ting Liu}, \bibinfo{person}{Daxin Jiang}, {et~al\mbox{.}}} \bibinfo{year}{2020}\natexlab{}.
\newblock \showarticletitle{Codebert: A pre-trained model for programming and natural languages}.
\newblock \bibinfo{journal}{\emph{arXiv preprint arXiv:2002.08155}} (\bibinfo{year}{2020}).
\newblock


\bibitem[Fu et~al\mbox{.}(2023)]%
        {fu2023effectiveness}
\bibfield{author}{\bibinfo{person}{Zihao Fu}, \bibinfo{person}{Haoran Yang}, \bibinfo{person}{Anthony Man-Cho So}, \bibinfo{person}{Wai Lam}, \bibinfo{person}{Lidong Bing}, {and} \bibinfo{person}{Nigel Collier}.} \bibinfo{year}{2023}\natexlab{}.
\newblock \showarticletitle{On the effectiveness of parameter-efficient fine-tuning}. In \bibinfo{booktitle}{\emph{Proceedings of the AAAI Conference on Artificial Intelligence}}, Vol.~\bibinfo{volume}{37}. \bibinfo{pages}{12799--12807}.
\newblock


\bibitem[Goel et~al\mbox{.}(2022)]%
        {goel2022cross}
\bibfield{author}{\bibinfo{person}{Divyam Goel}, \bibinfo{person}{Ramansh Grover}, {and} \bibinfo{person}{Fatemeh~H Fard}.} \bibinfo{year}{2022}\natexlab{}.
\newblock \showarticletitle{On the cross-modal transfer from natural language to code through adapter modules}. In \bibinfo{booktitle}{\emph{Proceedings of the 30th IEEE/ACM International Conference on Program Comprehension}}. \bibinfo{pages}{71--81}.
\newblock


\bibitem[Gou et~al\mbox{.}(2021)]%
        {gou2021knowledge}
\bibfield{author}{\bibinfo{person}{Jianping Gou}, \bibinfo{person}{Baosheng Yu}, \bibinfo{person}{Stephen~J Maybank}, {and} \bibinfo{person}{Dacheng Tao}.} \bibinfo{year}{2021}\natexlab{}.
\newblock \showarticletitle{Knowledge distillation: A survey}.
\newblock \bibinfo{journal}{\emph{International Journal of Computer Vision}}  \bibinfo{volume}{129} (\bibinfo{year}{2021}), \bibinfo{pages}{1789--1819}.
\newblock


\bibitem[Guo et~al\mbox{.}(2022)]%
        {guo2022unixcoder}
\bibfield{author}{\bibinfo{person}{Daya Guo}, \bibinfo{person}{Shuai Lu}, \bibinfo{person}{Nan Duan}, \bibinfo{person}{Yanlin Wang}, \bibinfo{person}{Ming Zhou}, {and} \bibinfo{person}{Jian Yin}.} \bibinfo{year}{2022}\natexlab{}.
\newblock \showarticletitle{Unixcoder: Unified cross-modal pre-training for code representation}.
\newblock \bibinfo{journal}{\emph{arXiv preprint arXiv:2203.03850}} (\bibinfo{year}{2022}).
\newblock


\bibitem[Guo et~al\mbox{.}(2020a)]%
        {guo2020graphcodebert}
\bibfield{author}{\bibinfo{person}{Daya Guo}, \bibinfo{person}{Shuo Ren}, \bibinfo{person}{Shuai Lu}, \bibinfo{person}{Zhangyin Feng}, \bibinfo{person}{Duyu Tang}, \bibinfo{person}{Shujie Liu}, \bibinfo{person}{Long Zhou}, \bibinfo{person}{Nan Duan}, \bibinfo{person}{Alexey Svyatkovskiy}, \bibinfo{person}{Shengyu Fu}, {et~al\mbox{.}}} \bibinfo{year}{2020}\natexlab{a}.
\newblock \showarticletitle{Graphcodebert: Pre-training code representations with data flow}.
\newblock \bibinfo{journal}{\emph{arXiv preprint arXiv:2009.08366}} (\bibinfo{year}{2020}).
\newblock


\bibitem[Guo et~al\mbox{.}(2020b)]%
        {guo2020parameter}
\bibfield{author}{\bibinfo{person}{Demi Guo}, \bibinfo{person}{Alexander~M Rush}, {and} \bibinfo{person}{Yoon Kim}.} \bibinfo{year}{2020}\natexlab{b}.
\newblock \showarticletitle{Parameter-efficient transfer learning with diff pruning}.
\newblock \bibinfo{journal}{\emph{arXiv preprint arXiv:2012.07463}} (\bibinfo{year}{2020}).
\newblock


\bibitem[Haque et~al\mbox{.}(2020)]%
        {haque2020improved}
\bibfield{author}{\bibinfo{person}{Sakib Haque}, \bibinfo{person}{Alexander LeClair}, \bibinfo{person}{Lingfei Wu}, {and} \bibinfo{person}{Collin McMillan}.} \bibinfo{year}{2020}\natexlab{}.
\newblock \showarticletitle{Improved automatic summarization of subroutines via attention to file context}. In \bibinfo{booktitle}{\emph{Proceedings of the 17th International Conference on Mining Software Repositories}}. \bibinfo{pages}{300--310}.
\newblock


\bibitem[He et~al\mbox{.}(2021b)]%
        {he2021towards}
\bibfield{author}{\bibinfo{person}{Junxian He}, \bibinfo{person}{Chunting Zhou}, \bibinfo{person}{Xuezhe Ma}, \bibinfo{person}{Taylor Berg-Kirkpatrick}, {and} \bibinfo{person}{Graham Neubig}.} \bibinfo{year}{2021}\natexlab{b}.
\newblock \showarticletitle{Towards a unified view of parameter-efficient transfer learning}.
\newblock \bibinfo{journal}{\emph{arXiv preprint arXiv:2110.04366}} (\bibinfo{year}{2021}).
\newblock


\bibitem[He et~al\mbox{.}(2021a)]%
        {he2021effectiveness}
\bibfield{author}{\bibinfo{person}{Ruidan He}, \bibinfo{person}{Linlin Liu}, \bibinfo{person}{Hai Ye}, \bibinfo{person}{Qingyu Tan}, \bibinfo{person}{Bosheng Ding}, \bibinfo{person}{Liying Cheng}, \bibinfo{person}{Jia-Wei Low}, \bibinfo{person}{Lidong Bing}, {and} \bibinfo{person}{Luo Si}.} \bibinfo{year}{2021}\natexlab{a}.
\newblock \showarticletitle{On the effectiveness of adapter-based tuning for pretrained language model adaptation}.
\newblock \bibinfo{journal}{\emph{arXiv preprint arXiv:2106.03164}} (\bibinfo{year}{2021}).
\newblock


\bibitem[Houlsby et~al\mbox{.}(2019)]%
        {houlsby2019parameter}
\bibfield{author}{\bibinfo{person}{Neil Houlsby}, \bibinfo{person}{Andrei Giurgiu}, \bibinfo{person}{Stanislaw Jastrzebski}, \bibinfo{person}{Bruna Morrone}, \bibinfo{person}{Quentin De~Laroussilhe}, \bibinfo{person}{Andrea Gesmundo}, \bibinfo{person}{Mona Attariyan}, {and} \bibinfo{person}{Sylvain Gelly}.} \bibinfo{year}{2019}\natexlab{}.
\newblock \showarticletitle{Parameter-efficient transfer learning for NLP}. In \bibinfo{booktitle}{\emph{International Conference on Machine Learning}}. PMLR, \bibinfo{pages}{2790--2799}.
\newblock


\bibitem[Hu et~al\mbox{.}(2021)]%
        {hu2021lora}
\bibfield{author}{\bibinfo{person}{Edward~J Hu}, \bibinfo{person}{Yelong Shen}, \bibinfo{person}{Phillip Wallis}, \bibinfo{person}{Zeyuan Allen-Zhu}, \bibinfo{person}{Yuanzhi Li}, \bibinfo{person}{Shean Wang}, \bibinfo{person}{Lu Wang}, {and} \bibinfo{person}{Weizhu Chen}.} \bibinfo{year}{2021}\natexlab{}.
\newblock \showarticletitle{Lora: Low-rank adaptation of large language models}.
\newblock \bibinfo{journal}{\emph{arXiv preprint arXiv:2106.09685}} (\bibinfo{year}{2021}).
\newblock


\bibitem[Husain et~al\mbox{.}(2019)]%
        {husain2019codesearchnet}
\bibfield{author}{\bibinfo{person}{Hamel Husain}, \bibinfo{person}{Ho-Hsiang Wu}, \bibinfo{person}{Tiferet Gazit}, \bibinfo{person}{Miltiadis Allamanis}, {and} \bibinfo{person}{Marc Brockschmidt}.} \bibinfo{year}{2019}\natexlab{}.
\newblock \showarticletitle{Codesearchnet challenge: Evaluating the state of semantic code search}.
\newblock \bibinfo{journal}{\emph{arXiv preprint arXiv:1909.09436}} (\bibinfo{year}{2019}).
\newblock


\bibitem[Iyer et~al\mbox{.}(2018)]%
        {iyer2018mapping}
\bibfield{author}{\bibinfo{person}{Srinivasan Iyer}, \bibinfo{person}{Ioannis Konstas}, \bibinfo{person}{Alvin Cheung}, {and} \bibinfo{person}{Luke Zettlemoyer}.} \bibinfo{year}{2018}\natexlab{}.
\newblock \showarticletitle{Mapping language to code in programmatic context}.
\newblock \bibinfo{journal}{\emph{arXiv preprint arXiv:1808.09588}} (\bibinfo{year}{2018}).
\newblock


\bibitem[Kanade et~al\mbox{.}(2020)]%
        {kanade2020learning}
\bibfield{author}{\bibinfo{person}{Aditya Kanade}, \bibinfo{person}{Petros Maniatis}, \bibinfo{person}{Gogul Balakrishnan}, {and} \bibinfo{person}{Kensen Shi}.} \bibinfo{year}{2020}\natexlab{}.
\newblock \showarticletitle{Learning and evaluating contextual embedding of source code}. In \bibinfo{booktitle}{\emph{International conference on machine learning}}. PMLR, \bibinfo{pages}{5110--5121}.
\newblock


\bibitem[Lester et~al\mbox{.}(2021)]%
        {lester2021power}
\bibfield{author}{\bibinfo{person}{Brian Lester}, \bibinfo{person}{Rami Al-Rfou}, {and} \bibinfo{person}{Noah Constant}.} \bibinfo{year}{2021}\natexlab{}.
\newblock \showarticletitle{The power of scale for parameter-efficient prompt tuning}.
\newblock \bibinfo{journal}{\emph{arXiv preprint arXiv:2104.08691}} (\bibinfo{year}{2021}).
\newblock


\bibitem[Lewis et~al\mbox{.}(2019)]%
        {lewis2019bart}
\bibfield{author}{\bibinfo{person}{Mike Lewis}, \bibinfo{person}{Yinhan Liu}, \bibinfo{person}{Naman Goyal}, \bibinfo{person}{Marjan Ghazvininejad}, \bibinfo{person}{Abdelrahman Mohamed}, \bibinfo{person}{Omer Levy}, \bibinfo{person}{Ves Stoyanov}, {and} \bibinfo{person}{Luke Zettlemoyer}.} \bibinfo{year}{2019}\natexlab{}.
\newblock \showarticletitle{Bart: Denoising sequence-to-sequence pre-training for natural language generation, translation, and comprehension}.
\newblock \bibinfo{journal}{\emph{arXiv preprint arXiv:1910.13461}} (\bibinfo{year}{2019}).
\newblock


\bibitem[Li and Liang(2021)]%
        {li2021prefix}
\bibfield{author}{\bibinfo{person}{Xiang~Lisa Li} {and} \bibinfo{person}{Percy Liang}.} \bibinfo{year}{2021}\natexlab{}.
\newblock \showarticletitle{Prefix-tuning: Optimizing continuous prompts for generation}.
\newblock \bibinfo{journal}{\emph{arXiv preprint arXiv:2101.00190}} (\bibinfo{year}{2021}).
\newblock


\bibitem[Lialin et~al\mbox{.}(2023)]%
        {lialin2023scaling}
\bibfield{author}{\bibinfo{person}{Vladislav Lialin}, \bibinfo{person}{Vijeta Deshpande}, {and} \bibinfo{person}{Anna Rumshisky}.} \bibinfo{year}{2023}\natexlab{}.
\newblock \showarticletitle{Scaling down to scale up: A guide to parameter-efficient fine-tuning}.
\newblock \bibinfo{journal}{\emph{arXiv preprint arXiv:2303.15647}} (\bibinfo{year}{2023}).
\newblock


\bibitem[Liu et~al\mbox{.}(2022)]%
        {liu2022few}
\bibfield{author}{\bibinfo{person}{Haokun Liu}, \bibinfo{person}{Derek Tam}, \bibinfo{person}{Mohammed Muqeeth}, \bibinfo{person}{Jay Mohta}, \bibinfo{person}{Tenghao Huang}, \bibinfo{person}{Mohit Bansal}, {and} \bibinfo{person}{Colin~A Raffel}.} \bibinfo{year}{2022}\natexlab{}.
\newblock \showarticletitle{Few-shot parameter-efficient fine-tuning is better and cheaper than in-context learning}.
\newblock \bibinfo{journal}{\emph{Advances in Neural Information Processing Systems}}  \bibinfo{volume}{35} (\bibinfo{year}{2022}), \bibinfo{pages}{1950--1965}.
\newblock


\bibitem[Liu et~al\mbox{.}(2023a)]%
        {liu2023empirical}
\bibfield{author}{\bibinfo{person}{Jiaxing Liu}, \bibinfo{person}{Chaofeng Sha}, {and} \bibinfo{person}{Xin Peng}.} \bibinfo{year}{2023}\natexlab{a}.
\newblock \showarticletitle{An Empirical Study of Parameter-Efficient Fine-Tuning Methods for Pre-trained Code Models}. In \bibinfo{booktitle}{\emph{2023 38th IEEE/ACM International Conference on Automated Software Engineering (ASE)}}. IEEE.
\newblock


\bibitem[Liu et~al\mbox{.}(2023b)]%
        {liu2023pre}
\bibfield{author}{\bibinfo{person}{Pengfei Liu}, \bibinfo{person}{Weizhe Yuan}, \bibinfo{person}{Jinlan Fu}, \bibinfo{person}{Zhengbao Jiang}, \bibinfo{person}{Hiroaki Hayashi}, {and} \bibinfo{person}{Graham Neubig}.} \bibinfo{year}{2023}\natexlab{b}.
\newblock \showarticletitle{Pre-train, prompt, and predict: A systematic survey of prompting methods in natural language processing}.
\newblock \bibinfo{journal}{\emph{Comput. Surveys}} \bibinfo{volume}{55}, \bibinfo{number}{9} (\bibinfo{year}{2023}), \bibinfo{pages}{1--35}.
\newblock


\bibitem[Liu et~al\mbox{.}(2023c)]%
        {liu2023gpt}
\bibfield{author}{\bibinfo{person}{Xiao Liu}, \bibinfo{person}{Yanan Zheng}, \bibinfo{person}{Zhengxiao Du}, \bibinfo{person}{Ming Ding}, \bibinfo{person}{Yujie Qian}, \bibinfo{person}{Zhilin Yang}, {and} \bibinfo{person}{Jie Tang}.} \bibinfo{year}{2023}\natexlab{c}.
\newblock \showarticletitle{GPT understands, too}.
\newblock \bibinfo{journal}{\emph{AI Open}} (\bibinfo{year}{2023}).
\newblock


\bibitem[Liu et~al\mbox{.}(2019)]%
        {liu2019roberta}
\bibfield{author}{\bibinfo{person}{Yinhan Liu}, \bibinfo{person}{Myle Ott}, \bibinfo{person}{Naman Goyal}, \bibinfo{person}{Jingfei Du}, \bibinfo{person}{Mandar Joshi}, \bibinfo{person}{Danqi Chen}, \bibinfo{person}{Omer Levy}, \bibinfo{person}{Mike Lewis}, \bibinfo{person}{Luke Zettlemoyer}, {and} \bibinfo{person}{Veselin Stoyanov}.} \bibinfo{year}{2019}\natexlab{}.
\newblock \showarticletitle{Roberta: A robustly optimized bert pretraining approach}.
\newblock \bibinfo{journal}{\emph{arXiv preprint arXiv:1907.11692}} (\bibinfo{year}{2019}).
\newblock


\bibitem[Liu et~al\mbox{.}(2021)]%
        {liu2021continual}
\bibfield{author}{\bibinfo{person}{Zihan Liu}, \bibinfo{person}{Genta~Indra Winata}, {and} \bibinfo{person}{Pascale Fung}.} \bibinfo{year}{2021}\natexlab{}.
\newblock \showarticletitle{Continual mixed-language pre-training for extremely low-resource neural machine translation}.
\newblock \bibinfo{journal}{\emph{arXiv preprint arXiv:2105.03953}} (\bibinfo{year}{2021}).
\newblock


\bibitem[Lu et~al\mbox{.}(2021)]%
        {lu2021codexglue}
\bibfield{author}{\bibinfo{person}{Shuai Lu}, \bibinfo{person}{Daya Guo}, \bibinfo{person}{Shuo Ren}, \bibinfo{person}{Junjie Huang}, \bibinfo{person}{Alexey Svyatkovskiy}, \bibinfo{person}{Ambrosio Blanco}, \bibinfo{person}{Colin Clement}, \bibinfo{person}{Dawn Drain}, \bibinfo{person}{Daxin Jiang}, \bibinfo{person}{Duyu Tang}, {et~al\mbox{.}}} \bibinfo{year}{2021}\natexlab{}.
\newblock \showarticletitle{Codexglue: A machine learning benchmark dataset for code understanding and generation}.
\newblock \bibinfo{journal}{\emph{arXiv preprint arXiv:2102.04664}} (\bibinfo{year}{2021}).
\newblock


\bibitem[Ma et~al\mbox{.}(2023)]%
        {ma2023scope}
\bibfield{author}{\bibinfo{person}{Wei Ma}, \bibinfo{person}{Shangqing Liu}, \bibinfo{person}{Wenhan Wang}, \bibinfo{person}{Qiang Hu}, \bibinfo{person}{Ye Liu}, \bibinfo{person}{Cen Zhang}, \bibinfo{person}{Liming Nie}, {and} \bibinfo{person}{Yang Liu}.} \bibinfo{year}{2023}\natexlab{}.
\newblock \showarticletitle{The Scope of ChatGPT in Software Engineering: A Thorough Investigation}.
\newblock \bibinfo{journal}{\emph{arXiv preprint arXiv:2305.12138}} (\bibinfo{year}{2023}).
\newblock


\bibitem[Mastropaolo et~al\mbox{.}(2023)]%
        {mastropaolo2023robustness}
\bibfield{author}{\bibinfo{person}{Antonio Mastropaolo}, \bibinfo{person}{Luca Pascarella}, \bibinfo{person}{Emanuela Guglielmi}, \bibinfo{person}{Matteo Ciniselli}, \bibinfo{person}{Simone Scalabrino}, \bibinfo{person}{Rocco Oliveto}, {and} \bibinfo{person}{Gabriele Bavota}.} \bibinfo{year}{2023}\natexlab{}.
\newblock \showarticletitle{On the robustness of code generation techniques: An empirical study on github copilot}.
\newblock \bibinfo{journal}{\emph{arXiv preprint arXiv:2302.00438}} (\bibinfo{year}{2023}).
\newblock


\bibitem[Mogadala et~al\mbox{.}(2020)]%
        {mogadala2020integrating}
\bibfield{author}{\bibinfo{person}{Aditya Mogadala}, \bibinfo{person}{Xiaoyu Shen}, {and} \bibinfo{person}{Dietrich Klakow}.} \bibinfo{year}{2020}\natexlab{}.
\newblock \showarticletitle{Integrating image captioning with rule-based entity masking}.
\newblock \bibinfo{journal}{\emph{arXiv preprint arXiv:2007.11690}} (\bibinfo{year}{2020}).
\newblock


\bibitem[Mou et~al\mbox{.}(2016)]%
        {mou2016convolutional}
\bibfield{author}{\bibinfo{person}{Lili Mou}, \bibinfo{person}{Ge Li}, \bibinfo{person}{Lu Zhang}, \bibinfo{person}{Tao Wang}, {and} \bibinfo{person}{Zhi Jin}.} \bibinfo{year}{2016}\natexlab{}.
\newblock \showarticletitle{Convolutional neural networks over tree structures for programming language processing}. In \bibinfo{booktitle}{\emph{Proceedings of the AAAI conference on artificial intelligence}}, Vol.~\bibinfo{volume}{30}.
\newblock


\bibitem[Nafi et~al\mbox{.}(2019)]%
        {nafi2019clcdsa}
\bibfield{author}{\bibinfo{person}{Kawser~Wazed Nafi}, \bibinfo{person}{Tonny~Shekha Kar}, \bibinfo{person}{Banani Roy}, \bibinfo{person}{Chanchal~K Roy}, {and} \bibinfo{person}{Kevin~A Schneider}.} \bibinfo{year}{2019}\natexlab{}.
\newblock \showarticletitle{Clcdsa: cross language code clone detection using syntactical features and api documentation}. In \bibinfo{booktitle}{\emph{2019 34th IEEE/ACM International Conference on Automated Software Engineering (ASE)}}. IEEE, \bibinfo{pages}{1026--1037}.
\newblock


\bibitem[Nijkamp et~al\mbox{.}(2022)]%
        {nijkamp2022codegen}
\bibfield{author}{\bibinfo{person}{Erik Nijkamp}, \bibinfo{person}{Bo Pang}, \bibinfo{person}{Hiroaki Hayashi}, \bibinfo{person}{Lifu Tu}, \bibinfo{person}{Huan Wang}, \bibinfo{person}{Yingbo Zhou}, \bibinfo{person}{Silvio Savarese}, {and} \bibinfo{person}{Caiming Xiong}.} \bibinfo{year}{2022}\natexlab{}.
\newblock \showarticletitle{Codegen: An open large language model for code with multi-turn program synthesis}.
\newblock \bibinfo{journal}{\emph{arXiv preprint arXiv:2203.13474}} (\bibinfo{year}{2022}).
\newblock


\bibitem[Niu et~al\mbox{.}(2022a)]%
        {niu2022deep}
\bibfield{author}{\bibinfo{person}{Changan Niu}, \bibinfo{person}{Chuanyi Li}, \bibinfo{person}{Bin Luo}, {and} \bibinfo{person}{Vincent Ng}.} \bibinfo{year}{2022}\natexlab{a}.
\newblock \showarticletitle{Deep learning meets software engineering: A survey on pre-trained models of source code}.
\newblock \bibinfo{journal}{\emph{arXiv preprint arXiv:2205.11739}} (\bibinfo{year}{2022}).
\newblock


\bibitem[Niu et~al\mbox{.}(2023)]%
        {niu2023empirical}
\bibfield{author}{\bibinfo{person}{Changan Niu}, \bibinfo{person}{Chuanyi Li}, \bibinfo{person}{Vincent Ng}, \bibinfo{person}{Dongxiao Chen}, \bibinfo{person}{Jidong Ge}, {and} \bibinfo{person}{Bin Luo}.} \bibinfo{year}{2023}\natexlab{}.
\newblock \showarticletitle{An Empirical Comparison of Pre-Trained Models of Source Code}.
\newblock \bibinfo{journal}{\emph{arXiv preprint arXiv:2302.04026}} (\bibinfo{year}{2023}).
\newblock


\bibitem[Niu et~al\mbox{.}(2022b)]%
        {niu2022spt}
\bibfield{author}{\bibinfo{person}{Changan Niu}, \bibinfo{person}{Chuanyi Li}, \bibinfo{person}{Vincent Ng}, \bibinfo{person}{Jidong Ge}, \bibinfo{person}{Liguo Huang}, {and} \bibinfo{person}{Bin Luo}.} \bibinfo{year}{2022}\natexlab{b}.
\newblock \showarticletitle{Spt-code: Sequence-to-sequence pre-training for learning the representation of source code}.
\newblock \bibinfo{journal}{\emph{arXiv preprint arXiv:2201.01549}} (\bibinfo{year}{2022}).
\newblock


\bibitem[Papineni et~al\mbox{.}(2002)]%
        {papineni2002bleu}
\bibfield{author}{\bibinfo{person}{Kishore Papineni}, \bibinfo{person}{Salim Roukos}, \bibinfo{person}{Todd Ward}, {and} \bibinfo{person}{Wei-Jing Zhu}.} \bibinfo{year}{2002}\natexlab{}.
\newblock \showarticletitle{Bleu: a method for automatic evaluation of machine translation}. In \bibinfo{booktitle}{\emph{Proceedings of the 40th annual meeting of the Association for Computational Linguistics}}. \bibinfo{pages}{311--318}.
\newblock


\bibitem[Pfeiffer et~al\mbox{.}(2020a)]%
        {pfeiffer2020AdapterHub}
\bibfield{author}{\bibinfo{person}{Jonas Pfeiffer}, \bibinfo{person}{Andreas R\"uckl\'{e}}, \bibinfo{person}{Clifton Poth}, \bibinfo{person}{Aishwarya Kamath}, \bibinfo{person}{Ivan Vuli\'{c}}, \bibinfo{person}{Sebastian Ruder}, \bibinfo{person}{Kyunghyun Cho}, {and} \bibinfo{person}{Iryna Gurevych}.} \bibinfo{year}{2020}\natexlab{a}.
\newblock \showarticletitle{AdapterHub: A Framework for Adapting Transformers}. In \bibinfo{booktitle}{\emph{Proceedings of the 2020 Conference on Empirical Methods in Natural Language Processing (EMNLP 2020): Systems Demonstrations}}. \bibinfo{publisher}{Association for Computational Linguistics}, \bibinfo{address}{Online}, \bibinfo{pages}{46--54}.
\newblock
\urldef\tempurl%
\url{https://www.aclweb.org/anthology/2020.emnlp-demos.7}
\showURL{%
\tempurl}


\bibitem[Pfeiffer et~al\mbox{.}(2020b)]%
        {pfeiffer2020mad}
\bibfield{author}{\bibinfo{person}{Jonas Pfeiffer}, \bibinfo{person}{Ivan Vuli{\'c}}, \bibinfo{person}{Iryna Gurevych}, {and} \bibinfo{person}{Sebastian Ruder}.} \bibinfo{year}{2020}\natexlab{b}.
\newblock \showarticletitle{Mad-x: An adapter-based framework for multi-task cross-lingual transfer}.
\newblock \bibinfo{journal}{\emph{arXiv preprint arXiv:2005.00052}} (\bibinfo{year}{2020}).
\newblock


\bibitem[Radford et~al\mbox{.}(2018)]%
        {radford2018improving}
\bibfield{author}{\bibinfo{person}{Alec Radford}, \bibinfo{person}{Karthik Narasimhan}, \bibinfo{person}{Tim Salimans}, \bibinfo{person}{Ilya Sutskever}, {et~al\mbox{.}}} \bibinfo{year}{2018}\natexlab{}.
\newblock \showarticletitle{Improving language understanding by generative pre-training}.
\newblock  (\bibinfo{year}{2018}).
\newblock


\bibitem[Radford et~al\mbox{.}(2019)]%
        {radford2019language}
\bibfield{author}{\bibinfo{person}{Alec Radford}, \bibinfo{person}{Jeffrey Wu}, \bibinfo{person}{Rewon Child}, \bibinfo{person}{David Luan}, \bibinfo{person}{Dario Amodei}, \bibinfo{person}{Ilya Sutskever}, {et~al\mbox{.}}} \bibinfo{year}{2019}\natexlab{}.
\newblock \showarticletitle{Language models are unsupervised multitask learners}.
\newblock \bibinfo{journal}{\emph{OpenAI blog}} \bibinfo{volume}{1}, \bibinfo{number}{8} (\bibinfo{year}{2019}), \bibinfo{pages}{9}.
\newblock


\bibitem[Raffel et~al\mbox{.}(2020)]%
        {raffel2020exploring}
\bibfield{author}{\bibinfo{person}{Colin Raffel}, \bibinfo{person}{Noam Shazeer}, \bibinfo{person}{Adam Roberts}, \bibinfo{person}{Katherine Lee}, \bibinfo{person}{Sharan Narang}, \bibinfo{person}{Michael Matena}, \bibinfo{person}{Yanqi Zhou}, \bibinfo{person}{Wei Li}, {and} \bibinfo{person}{Peter~J Liu}.} \bibinfo{year}{2020}\natexlab{}.
\newblock \showarticletitle{Exploring the limits of transfer learning with a unified text-to-text transformer}.
\newblock \bibinfo{journal}{\emph{The Journal of Machine Learning Research}} \bibinfo{volume}{21}, \bibinfo{number}{1} (\bibinfo{year}{2020}), \bibinfo{pages}{5485--5551}.
\newblock


\bibitem[Rebuffi et~al\mbox{.}(2017)]%
        {rebuffi2017learning}
\bibfield{author}{\bibinfo{person}{Sylvestre-Alvise Rebuffi}, \bibinfo{person}{Hakan Bilen}, {and} \bibinfo{person}{Andrea Vedaldi}.} \bibinfo{year}{2017}\natexlab{}.
\newblock \showarticletitle{Learning multiple visual domains with residual adapters}.
\newblock \bibinfo{journal}{\emph{Advances in neural information processing systems}}  \bibinfo{volume}{30} (\bibinfo{year}{2017}).
\newblock


\bibitem[Ren et~al\mbox{.}(2020)]%
        {ren2020codebleu}
\bibfield{author}{\bibinfo{person}{Shuo Ren}, \bibinfo{person}{Daya Guo}, \bibinfo{person}{Shuai Lu}, \bibinfo{person}{Long Zhou}, \bibinfo{person}{Shujie Liu}, \bibinfo{person}{Duyu Tang}, \bibinfo{person}{Neel Sundaresan}, \bibinfo{person}{Ming Zhou}, \bibinfo{person}{Ambrosio Blanco}, {and} \bibinfo{person}{Shuai Ma}.} \bibinfo{year}{2020}\natexlab{}.
\newblock \showarticletitle{Codebleu: a method for automatic evaluation of code synthesis}.
\newblock \bibinfo{journal}{\emph{arXiv preprint arXiv:2009.10297}} (\bibinfo{year}{2020}).
\newblock


\bibitem[Roziere et~al\mbox{.}(2023)]%
        {roziere2023code}
\bibfield{author}{\bibinfo{person}{Baptiste Roziere}, \bibinfo{person}{Jonas Gehring}, \bibinfo{person}{Fabian Gloeckle}, \bibinfo{person}{Sten Sootla}, \bibinfo{person}{Itai Gat}, \bibinfo{person}{Xiaoqing~Ellen Tan}, \bibinfo{person}{Yossi Adi}, \bibinfo{person}{Jingyu Liu}, \bibinfo{person}{Tal Remez}, \bibinfo{person}{J{\'e}r{\'e}my Rapin}, {et~al\mbox{.}}} \bibinfo{year}{2023}\natexlab{}.
\newblock \showarticletitle{Code llama: Open foundation models for code}.
\newblock \bibinfo{journal}{\emph{arXiv preprint arXiv:2308.12950}} (\bibinfo{year}{2023}).
\newblock


\bibitem[Saberi and Fard(2023)]%
        {saberi2023model}
\bibfield{author}{\bibinfo{person}{Iman Saberi} {and} \bibinfo{person}{Fatemeh~H Fard}.} \bibinfo{year}{2023}\natexlab{}.
\newblock \showarticletitle{Model-Agnostic Syntactical Information for Pre-Trained Programming Language Models}.
\newblock \bibinfo{journal}{\emph{arXiv preprint arXiv:2303.06233}} (\bibinfo{year}{2023}).
\newblock


\bibitem[Shen et~al\mbox{.}(2022)]%
        {shen2022semipqa}
\bibfield{author}{\bibinfo{person}{Xiaoyu Shen}, \bibinfo{person}{Gianni Barlacchi}, \bibinfo{person}{Marco Del~Tredici}, \bibinfo{person}{Weiwei Cheng}, {and} \bibinfo{person}{Adri{\`a} de Gispert}.} \bibinfo{year}{2022}\natexlab{}.
\newblock \showarticletitle{semiPQA: A Study on Product Question Answering over Semi-structured Data}. In \bibinfo{booktitle}{\emph{Proceedings of the Fifth Workshop on e-Commerce and NLP (ECNLP 5)}}. \bibinfo{pages}{111--120}.
\newblock


\bibitem[Shen et~al\mbox{.}(2017)]%
        {shen2017estimation}
\bibfield{author}{\bibinfo{person}{Xiaoyu Shen}, \bibinfo{person}{Youssef Oualil}, \bibinfo{person}{Clayton Greenberg}, \bibinfo{person}{Mittul Singh}, {and} \bibinfo{person}{Dietrich Klakow}.} \bibinfo{year}{2017}\natexlab{}.
\newblock \showarticletitle{Estimation of Gap Between Current Language Models and Human Performance.}
\newblock


\bibitem[Shi et~al\mbox{.}(2023)]%
        {shi2023towards}
\bibfield{author}{\bibinfo{person}{Ensheng Shi}, \bibinfo{person}{Yanlin Wang}, \bibinfo{person}{Hongyu Zhang}, \bibinfo{person}{Lun Du}, \bibinfo{person}{Shi Han}, \bibinfo{person}{Dongmei Zhang}, {and} \bibinfo{person}{Hongbin Sun}.} \bibinfo{year}{2023}\natexlab{}.
\newblock \showarticletitle{Towards Efficient Fine-tuning of Pre-trained Code Models: An Experimental Study and Beyond}.
\newblock \bibinfo{journal}{\emph{arXiv preprint arXiv:2304.05216}} (\bibinfo{year}{2023}).
\newblock


\bibitem[Shi et~al\mbox{.}(2022)]%
        {shi2022compressing}
\bibfield{author}{\bibinfo{person}{Jieke Shi}, \bibinfo{person}{Zhou Yang}, \bibinfo{person}{Bowen Xu}, \bibinfo{person}{Hong~Jin Kang}, {and} \bibinfo{person}{David Lo}.} \bibinfo{year}{2022}\natexlab{}.
\newblock \showarticletitle{Compressing pre-trained models of code into 3 mb}. In \bibinfo{booktitle}{\emph{Proceedings of the 37th IEEE/ACM International Conference on Automated Software Engineering}}. \bibinfo{pages}{1--12}.
\newblock


\bibitem[Song et~al\mbox{.}(2022)]%
        {song2022multi}
\bibfield{author}{\bibinfo{person}{Dezhao Song}, \bibinfo{person}{Andrew Vold}, \bibinfo{person}{Kanika Madan}, {and} \bibinfo{person}{Frank Schilder}.} \bibinfo{year}{2022}\natexlab{}.
\newblock \showarticletitle{Multi-label legal document classification: A deep learning-based approach with label-attention and domain-specific pre-training}.
\newblock \bibinfo{journal}{\emph{Information Systems}}  \bibinfo{volume}{106} (\bibinfo{year}{2022}), \bibinfo{pages}{101718}.
\newblock


\bibitem[Steenhoek et~al\mbox{.}(2022)]%
        {steenhoek2022empirical}
\bibfield{author}{\bibinfo{person}{Benjamin Steenhoek}, \bibinfo{person}{Md~Mahbubur Rahman}, \bibinfo{person}{Richard Jiles}, {and} \bibinfo{person}{Wei Le}.} \bibinfo{year}{2022}\natexlab{}.
\newblock \showarticletitle{An Empirical Study of Deep Learning Models for Vulnerability Detection}.
\newblock \bibinfo{journal}{\emph{arXiv preprint arXiv:2212.08109}} (\bibinfo{year}{2022}).
\newblock


\bibitem[Steenhoek et~al\mbox{.}(2023)]%
        {steenhoek2023empirical}
\bibfield{author}{\bibinfo{person}{Benjamin Steenhoek}, \bibinfo{person}{Md~Mahbubur Rahman}, \bibinfo{person}{Richard Jiles}, {and} \bibinfo{person}{Wei Le}.} \bibinfo{year}{2023}\natexlab{}.
\newblock \showarticletitle{An empirical study of deep learning models for vulnerability detection}. In \bibinfo{booktitle}{\emph{2023 IEEE/ACM 45th International Conference on Software Engineering (ICSE)}}. IEEE, \bibinfo{pages}{2237--2248}.
\newblock


\bibitem[Su et~al\mbox{.}(2020)]%
        {su2020moviechats}
\bibfield{author}{\bibinfo{person}{Hui Su}, \bibinfo{person}{Xiaoyu Shen}, \bibinfo{person}{Zhou Xiao}, \bibinfo{person}{Zheng Zhang}, \bibinfo{person}{Ernie Chang}, \bibinfo{person}{Cheng Zhang}, \bibinfo{person}{Cheng Niu}, {and} \bibinfo{person}{Jie Zhou}.} \bibinfo{year}{2020}\natexlab{}.
\newblock \showarticletitle{Moviechats: Chat like humans in a closed domain}. In \bibinfo{booktitle}{\emph{Proceedings of the 2020 conference on empirical methods in natural language processing (EMNLP)}}. \bibinfo{pages}{6605--6619}.
\newblock


\bibitem[Su et~al\mbox{.}(2022a)]%
        {su2022rocbert}
\bibfield{author}{\bibinfo{person}{Hui Su}, \bibinfo{person}{Weiwei Shi}, \bibinfo{person}{Xiaoyu Shen}, \bibinfo{person}{Zhou Xiao}, \bibinfo{person}{Tuo Ji}, \bibinfo{person}{Jiarui Fang}, {and} \bibinfo{person}{Jie Zhou}.} \bibinfo{year}{2022}\natexlab{a}.
\newblock \showarticletitle{Rocbert: Robust chinese bert with multimodal contrastive pretraining}. In \bibinfo{booktitle}{\emph{Proceedings of the 60th Annual Meeting of the Association for Computational Linguistics (Volume 1: Long Papers)}}. \bibinfo{pages}{921--931}.
\newblock


\bibitem[Su et~al\mbox{.}(2022b)]%
        {su2022welm}
\bibfield{author}{\bibinfo{person}{Hui Su}, \bibinfo{person}{Xiao Zhou}, \bibinfo{person}{Houjin Yu}, \bibinfo{person}{Xiaoyu Shen}, \bibinfo{person}{Yuwen Chen}, \bibinfo{person}{Zilin Zhu}, \bibinfo{person}{Yang Yu}, {and} \bibinfo{person}{Jie Zhou}.} \bibinfo{year}{2022}\natexlab{b}.
\newblock \showarticletitle{Welm: A well-read pre-trained language model for chinese}.
\newblock \bibinfo{journal}{\emph{arXiv preprint arXiv:2209.10372}} (\bibinfo{year}{2022}).
\newblock


\bibitem[Svajlenko et~al\mbox{.}(2014)]%
        {svajlenko2014towards}
\bibfield{author}{\bibinfo{person}{Jeffrey Svajlenko}, \bibinfo{person}{Judith~F Islam}, \bibinfo{person}{Iman Keivanloo}, \bibinfo{person}{Chanchal~K Roy}, {and} \bibinfo{person}{Mohammad~Mamun Mia}.} \bibinfo{year}{2014}\natexlab{}.
\newblock \showarticletitle{Towards a big data curated benchmark of inter-project code clones}. In \bibinfo{booktitle}{\emph{2014 IEEE International Conference on Software Maintenance and Evolution}}. IEEE, \bibinfo{pages}{476--480}.
\newblock


\bibitem[Svyatkovskiy et~al\mbox{.}(2020)]%
        {svyatkovskiy2020intellicode}
\bibfield{author}{\bibinfo{person}{Alexey Svyatkovskiy}, \bibinfo{person}{Shao~Kun Deng}, \bibinfo{person}{Shengyu Fu}, {and} \bibinfo{person}{Neel Sundaresan}.} \bibinfo{year}{2020}\natexlab{}.
\newblock \showarticletitle{Intellicode compose: Code generation using transformer}. In \bibinfo{booktitle}{\emph{Proceedings of the 28th ACM Joint Meeting on European Software Engineering Conference and Symposium on the Foundations of Software Engineering}}. \bibinfo{pages}{1433--1443}.
\newblock


\bibitem[Svyatkovskiy et~al\mbox{.}(2021)]%
        {svyatkovskiy2021fast}
\bibfield{author}{\bibinfo{person}{Alexey Svyatkovskiy}, \bibinfo{person}{Sebastian Lee}, \bibinfo{person}{Anna Hadjitofi}, \bibinfo{person}{Maik Riechert}, \bibinfo{person}{Juliana~Vicente Franco}, {and} \bibinfo{person}{Miltiadis Allamanis}.} \bibinfo{year}{2021}\natexlab{}.
\newblock \showarticletitle{Fast and memory-efficient neural code completion}. In \bibinfo{booktitle}{\emph{2021 IEEE/ACM 18th International Conference on Mining Software Repositories (MSR)}}. IEEE, \bibinfo{pages}{329--340}.
\newblock


\bibitem[Tang et~al\mbox{.}(2021)]%
        {tang2021ast}
\bibfield{author}{\bibinfo{person}{Ze Tang}, \bibinfo{person}{Chuanyi Li}, \bibinfo{person}{Jidong Ge}, \bibinfo{person}{Xiaoyu Shen}, \bibinfo{person}{Zheling Zhu}, {and} \bibinfo{person}{Bin Luo}.} \bibinfo{year}{2021}\natexlab{}.
\newblock \showarticletitle{AST-transformer: Encoding abstract syntax trees efficiently for code summarization}. In \bibinfo{booktitle}{\emph{2021 36th IEEE/ACM International Conference on Automated Software Engineering (ASE)}}. IEEE, \bibinfo{pages}{1193--1195}.
\newblock


\bibitem[Tang et~al\mbox{.}(2022)]%
        {tang2022ast}
\bibfield{author}{\bibinfo{person}{Ze Tang}, \bibinfo{person}{Xiaoyu Shen}, \bibinfo{person}{Chuanyi Li}, \bibinfo{person}{Jidong Ge}, \bibinfo{person}{Liguo Huang}, \bibinfo{person}{Zhelin Zhu}, {and} \bibinfo{person}{Bin Luo}.} \bibinfo{year}{2022}\natexlab{}.
\newblock \showarticletitle{AST-trans: Code summarization with efficient tree-structured attention}. In \bibinfo{booktitle}{\emph{Proceedings of the 44th International Conference on Software Engineering}}. \bibinfo{pages}{150--162}.
\newblock


\bibitem[Tufano et~al\mbox{.}(2023)]%
        {tufano2023automating}
\bibfield{author}{\bibinfo{person}{Rosalia Tufano}, \bibinfo{person}{Luca Pascarella}, {and} \bibinfo{person}{Gabriele Bavota}.} \bibinfo{year}{2023}\natexlab{}.
\newblock \showarticletitle{Automating Code-Related Tasks Through Transformers: The Impact of Pre-training}.
\newblock \bibinfo{journal}{\emph{arXiv preprint arXiv:2302.04048}} (\bibinfo{year}{2023}).
\newblock


\bibitem[Vaswani et~al\mbox{.}(2017)]%
        {vaswani2017attention}
\bibfield{author}{\bibinfo{person}{Ashish Vaswani}, \bibinfo{person}{Noam Shazeer}, \bibinfo{person}{Niki Parmar}, \bibinfo{person}{Jakob Uszkoreit}, \bibinfo{person}{Llion Jones}, \bibinfo{person}{Aidan~N Gomez}, \bibinfo{person}{{\L}ukasz Kaiser}, {and} \bibinfo{person}{Illia Polosukhin}.} \bibinfo{year}{2017}\natexlab{}.
\newblock \showarticletitle{Attention is all you need}.
\newblock \bibinfo{journal}{\emph{Advances in neural information processing systems}}  \bibinfo{volume}{30} (\bibinfo{year}{2017}).
\newblock


\bibitem[Wang et~al\mbox{.}(2022b)]%
        {wang2022no}
\bibfield{author}{\bibinfo{person}{Chaozheng Wang}, \bibinfo{person}{Yuanhang Yang}, \bibinfo{person}{Cuiyun Gao}, \bibinfo{person}{Yun Peng}, \bibinfo{person}{Hongyu Zhang}, {and} \bibinfo{person}{Michael~R Lyu}.} \bibinfo{year}{2022}\natexlab{b}.
\newblock \showarticletitle{No more fine-tuning? an experimental evaluation of prompt tuning in code intelligence}. In \bibinfo{booktitle}{\emph{Proceedings of the 30th ACM Joint European Software Engineering Conference and Symposium on the Foundations of Software Engineering}}. \bibinfo{pages}{382--394}.
\newblock


\bibitem[Wang et~al\mbox{.}(2023)]%
        {wang2023one}
\bibfield{author}{\bibinfo{person}{Deze Wang}, \bibinfo{person}{Boxing Chen}, \bibinfo{person}{Shanshan Li}, \bibinfo{person}{Wei Luo}, \bibinfo{person}{Shaoliang Peng}, \bibinfo{person}{Wei Dong}, {and} \bibinfo{person}{Xiangke Liao}.} \bibinfo{year}{2023}\natexlab{}.
\newblock \showarticletitle{One Adapter for All Programming Languages? Adapter Tuning for Code Search and Summarization}.
\newblock \bibinfo{journal}{\emph{arXiv preprint arXiv:2303.15822}} (\bibinfo{year}{2023}).
\newblock


\bibitem[Wang et~al\mbox{.}(2022a)]%
        {wang2022bridging}
\bibfield{author}{\bibinfo{person}{Deze Wang}, \bibinfo{person}{Zhouyang Jia}, \bibinfo{person}{Shanshan Li}, \bibinfo{person}{Yue Yu}, \bibinfo{person}{Yun Xiong}, \bibinfo{person}{Wei Dong}, {and} \bibinfo{person}{Xiangke Liao}.} \bibinfo{year}{2022}\natexlab{a}.
\newblock \showarticletitle{Bridging pre-trained models and downstream tasks for source code understanding}. In \bibinfo{booktitle}{\emph{Proceedings of the 44th International Conference on Software Engineering}}. \bibinfo{pages}{287--298}.
\newblock


\bibitem[Wang et~al\mbox{.}(2021)]%
        {wang2021codet5}
\bibfield{author}{\bibinfo{person}{Yue Wang}, \bibinfo{person}{Weishi Wang}, \bibinfo{person}{Shafiq Joty}, {and} \bibinfo{person}{Steven~CH Hoi}.} \bibinfo{year}{2021}\natexlab{}.
\newblock \showarticletitle{Codet5: Identifier-aware unified pre-trained encoder-decoder models for code understanding and generation}.
\newblock \bibinfo{journal}{\emph{arXiv preprint arXiv:2109.00859}} (\bibinfo{year}{2021}).
\newblock


\bibitem[Weiss et~al\mbox{.}(2016)]%
        {weiss2016survey}
\bibfield{author}{\bibinfo{person}{Karl Weiss}, \bibinfo{person}{Taghi~M Khoshgoftaar}, {and} \bibinfo{person}{DingDing Wang}.} \bibinfo{year}{2016}\natexlab{}.
\newblock \showarticletitle{A survey of transfer learning}.
\newblock \bibinfo{journal}{\emph{Journal of Big data}} \bibinfo{volume}{3}, \bibinfo{number}{1} (\bibinfo{year}{2016}), \bibinfo{pages}{1--40}.
\newblock


\bibitem[Xu et~al\mbox{.}(2019)]%
        {xu2019bert}
\bibfield{author}{\bibinfo{person}{Hu Xu}, \bibinfo{person}{Bing Liu}, \bibinfo{person}{Lei Shu}, {and} \bibinfo{person}{Philip~S Yu}.} \bibinfo{year}{2019}\natexlab{}.
\newblock \showarticletitle{BERT post-training for review reading comprehension and aspect-based sentiment analysis}.
\newblock \bibinfo{journal}{\emph{arXiv preprint arXiv:1904.02232}} (\bibinfo{year}{2019}).
\newblock


\bibitem[Yan et~al\mbox{.}(2020)]%
        {yan2020code}
\bibfield{author}{\bibinfo{person}{Shuhan Yan}, \bibinfo{person}{Hang Yu}, \bibinfo{person}{Yuting Chen}, \bibinfo{person}{Beijun Shen}, {and} \bibinfo{person}{Lingxiao Jiang}.} \bibinfo{year}{2020}\natexlab{}.
\newblock \showarticletitle{Are the code snippets what we are searching for? a benchmark and an empirical study on code search with natural-language queries}. In \bibinfo{booktitle}{\emph{2020 IEEE 27th International Conference on Software Analysis, Evolution and Reengineering (SANER)}}. IEEE, \bibinfo{pages}{344--354}.
\newblock


\bibitem[Yang et~al\mbox{.}(2023)]%
        {yang2023deep}
\bibfield{author}{\bibinfo{person}{Zezhou Yang}, \bibinfo{person}{Sirong Chen}, \bibinfo{person}{Cuiyun Gao}, \bibinfo{person}{Zhenhao Li}, \bibinfo{person}{Ge Li}, {and} \bibinfo{person}{Rongcong Lv}.} \bibinfo{year}{2023}\natexlab{}.
\newblock \showarticletitle{Deep Learning Based Code Generation Methods: A Literature Review}.
\newblock \bibinfo{journal}{\emph{arXiv preprint arXiv:2303.01056}} (\bibinfo{year}{2023}).
\newblock


\bibitem[Yang et~al\mbox{.}(2019)]%
        {yang2019xlnet}
\bibfield{author}{\bibinfo{person}{Zhilin Yang}, \bibinfo{person}{Zihang Dai}, \bibinfo{person}{Yiming Yang}, \bibinfo{person}{Jaime Carbonell}, \bibinfo{person}{Russ~R Salakhutdinov}, {and} \bibinfo{person}{Quoc~V Le}.} \bibinfo{year}{2019}\natexlab{}.
\newblock \showarticletitle{Xlnet: Generalized autoregressive pretraining for language understanding}.
\newblock \bibinfo{journal}{\emph{Advances in neural information processing systems}}  \bibinfo{volume}{32} (\bibinfo{year}{2019}).
\newblock


\bibitem[Zaken et~al\mbox{.}(2021)]%
        {zaken2021bitfit}
\bibfield{author}{\bibinfo{person}{Elad~Ben Zaken}, \bibinfo{person}{Shauli Ravfogel}, {and} \bibinfo{person}{Yoav Goldberg}.} \bibinfo{year}{2021}\natexlab{}.
\newblock \showarticletitle{Bitfit: Simple parameter-efficient fine-tuning for transformer-based masked language-models}.
\newblock \bibinfo{journal}{\emph{arXiv preprint arXiv:2106.10199}} (\bibinfo{year}{2021}).
\newblock


\bibitem[Zan et~al\mbox{.}(2023)]%
        {zan2023large}
\bibfield{author}{\bibinfo{person}{Daoguang Zan}, \bibinfo{person}{Bei Chen}, \bibinfo{person}{Fengji Zhang}, \bibinfo{person}{Dianjie Lu}, \bibinfo{person}{Bingchao Wu}, \bibinfo{person}{Bei Guan}, \bibinfo{person}{Wang Yongji}, {and} \bibinfo{person}{Jian-Guang Lou}.} \bibinfo{year}{2023}\natexlab{}.
\newblock \showarticletitle{Large language models meet NL2Code: A survey}. In \bibinfo{booktitle}{\emph{Proceedings of the 61st Annual Meeting of the Association for Computational Linguistics (Volume 1: Long Papers)}}. \bibinfo{pages}{7443--7464}.
\newblock


\bibitem[Zeng et~al\mbox{.}(2022)]%
        {zeng2022extensive}
\bibfield{author}{\bibinfo{person}{Zhengran Zeng}, \bibinfo{person}{Hanzhuo Tan}, \bibinfo{person}{Haotian Zhang}, \bibinfo{person}{Jing Li}, \bibinfo{person}{Yuqun Zhang}, {and} \bibinfo{person}{Lingming Zhang}.} \bibinfo{year}{2022}\natexlab{}.
\newblock \showarticletitle{An extensive study on pre-trained models for program understanding and generation}. In \bibinfo{booktitle}{\emph{Proceedings of the 31st ACM SIGSOFT International Symposium on Software Testing and Analysis}}. \bibinfo{pages}{39--51}.
\newblock


\bibitem[Zhang et~al\mbox{.}(2023)]%
        {zhang2023adaptive}
\bibfield{author}{\bibinfo{person}{Qingru Zhang}, \bibinfo{person}{Minshuo Chen}, \bibinfo{person}{Alexander Bukharin}, \bibinfo{person}{Pengcheng He}, \bibinfo{person}{Yu Cheng}, \bibinfo{person}{Weizhu Chen}, {and} \bibinfo{person}{Tuo Zhao}.} \bibinfo{year}{2023}\natexlab{}.
\newblock \showarticletitle{Adaptive budget allocation for parameter-efficient fine-tuning}.
\newblock \bibinfo{journal}{\emph{arXiv preprint arXiv:2303.10512}} (\bibinfo{year}{2023}).
\newblock


\bibitem[Zhang et~al\mbox{.}(2022)]%
        {zhang2022mdia}
\bibfield{author}{\bibinfo{person}{Qingyu Zhang}, \bibinfo{person}{Xiaoyu Shen}, \bibinfo{person}{Ernie Chang}, \bibinfo{person}{Jidong Ge}, {and} \bibinfo{person}{Pengke Chen}.} \bibinfo{year}{2022}\natexlab{}.
\newblock \showarticletitle{Mdia: A benchmark for multilingual dialogue generation in 46 languages}.
\newblock \bibinfo{journal}{\emph{arXiv preprint arXiv:2208.13078}} (\bibinfo{year}{2022}).
\newblock


\bibitem[Zhang and Yang(2018)]%
        {zhang2018overview}
\bibfield{author}{\bibinfo{person}{Yu Zhang} {and} \bibinfo{person}{Qiang Yang}.} \bibinfo{year}{2018}\natexlab{}.
\newblock \showarticletitle{An overview of multi-task learning}.
\newblock \bibinfo{journal}{\emph{National Science Review}} \bibinfo{volume}{5}, \bibinfo{number}{1} (\bibinfo{year}{2018}), \bibinfo{pages}{30--43}.
\newblock


\bibitem[Zheng et~al\mbox{.}(2023)]%
        {zheng2023towards}
\bibfield{author}{\bibinfo{person}{Zibin Zheng}, \bibinfo{person}{Kaiwen Ning}, \bibinfo{person}{Jiachi Chen}, \bibinfo{person}{Yanlin Wang}, \bibinfo{person}{Wenqing Chen}, \bibinfo{person}{Lianghong Guo}, {and} \bibinfo{person}{Weicheng Wang}.} \bibinfo{year}{2023}\natexlab{}.
\newblock \showarticletitle{Towards an understanding of large language models in software engineering tasks}.
\newblock \bibinfo{journal}{\emph{arXiv preprint arXiv:2308.11396}} (\bibinfo{year}{2023}).
\newblock


\bibitem[Zhong et~al\mbox{.}(2022)]%
        {zhong2022standup4npr}
\bibfield{author}{\bibinfo{person}{Wenkang Zhong}, \bibinfo{person}{Hongliang Ge}, \bibinfo{person}{Hongfei Ai}, \bibinfo{person}{Chuanyi Li}, \bibinfo{person}{Kui Liu}, \bibinfo{person}{Jidong Ge}, {and} \bibinfo{person}{Bin Luo}.} \bibinfo{year}{2022}\natexlab{}.
\newblock \showarticletitle{StandUp4NPR: Standardizing SetUp for Empirically Comparing Neural Program Repair Systems}. In \bibinfo{booktitle}{\emph{37th IEEE/ACM International Conference on Automated Software Engineering}}. \bibinfo{pages}{1--13}.
\newblock


\bibitem[Zhou et~al\mbox{.}(2021)]%
        {zhou2021assessing}
\bibfield{author}{\bibinfo{person}{Xin Zhou}, \bibinfo{person}{DongGyun Han}, {and} \bibinfo{person}{David Lo}.} \bibinfo{year}{2021}\natexlab{}.
\newblock \showarticletitle{Assessing generalizability of codebert}. In \bibinfo{booktitle}{\emph{2021 IEEE International Conference on Software Maintenance and Evolution (ICSME)}}. IEEE, \bibinfo{pages}{425--436}.
\newblock


\bibitem[Zhou et~al\mbox{.}(2019)]%
        {zhou2019devign}
\bibfield{author}{\bibinfo{person}{Yaqin Zhou}, \bibinfo{person}{Shangqing Liu}, \bibinfo{person}{Jingkai Siow}, \bibinfo{person}{Xiaoning Du}, {and} \bibinfo{person}{Yang Liu}.} \bibinfo{year}{2019}\natexlab{}.
\newblock \showarticletitle{Devign: Effective vulnerability identification by learning comprehensive program semantics via graph neural networks}.
\newblock \bibinfo{journal}{\emph{Advances in neural information processing systems}}  \bibinfo{volume}{32} (\bibinfo{year}{2019}).
\newblock


\end{thebibliography}

\appendix

\end{document}